\documentclass[sigconf]{acmart}

\AtBeginDocument{%
  }

\setcopyright{none}
\copyrightyear{2026}
\acmYear{2026}
\acmConference{Academia x Industry for Silicon Design}{2026}{}
\renewcommand\footnotetextcopyrightpermission[1]{}
\usepackage{subcaption}
\usepackage{listings}
\usepackage{flushend}
\usepackage{array}
\usepackage{multirow}

\lstdefinelanguage{yaml}{
  keywords={true, false, null, yes, no},
  keywordstyle=\color{blue}\bfseries,
  basicstyle=\ttfamily\small,
  sensitive=false,
  comment=[l]{\#},
  commentstyle=\color{gray}\ttfamily,
  stringstyle=\color{red},
  morestring=[b]',
  morestring=[b]",
  moredelim=[l][\color{orange}]{-\ },
  moredelim=**[l][\color{teal}]{:\ },
}
\usepackage{cleveref}

\usepackage{xspace}


\begin{document}

\title{Academia x Industry: The Role of Fundamentals for Silicon in an AI Native Era}

\author{Vincent T. Lee}
\affiliation{%
  \institution{Meta Reality Labs Silicon}
  \city{Seattle}
  \state{WA}
  \country{USA}
}
\email{vtlee@meta.com}

\author{Armin Alaghi}
\affiliation{%
  \institution{Meta Reality Labs Research}
  \city{Redmond}
  \state{WA}
  \country{USA}
}
\email{alaghi@meta.com}

\author{Carole-Jean Wu}
\affiliation{%
  \institution{Meta FAIR}
  \city{Cambridge}
  \state{MA}
  \country{USA}
}
\email{carolejeanwu@meta.com}

\author{Sai Zhang}
\affiliation{%
  \institution{New York University}
  \city{New York}
  \state{NY}
  \country{USA}
}
\email{sai.zhang@nyu.edu}

\author{Brandon Reagen}
\affiliation{%
  \institution{New York University}
  \city{New York}
  \state{NY}
  \country{USA}
}
\email{bjr5@nyu.edu}

\author{Thierry Tambe}
\affiliation{%
  \institution{Stanford University}
  \city{Stanford}
  \state{CA}
  \country{USA}
}
\email{ttambe@stanford.edu}

\author{Jean Boufarhat}
\affiliation{%
  \institution{Meta Reality Labs Silicon}
  \city{Sunnyvale}
  \state{CA}
  \country{USA}
}
\email{jeanboufarhat@meta.com}

\author{Matheus Trevisan Moreira}
\affiliation{%
  \institution{Meta Reality Labs Silicon}
  \city{San Diego}
  \state{CA}
  \country{USA}
}
\email{matheustrev@meta.com}

\begin{abstract}
Agentic AI is set to become one of the most transformational technologies in generations and materially change how we approach silicon design and engineering. The impact is being felt in real time amid a rapidly changing landscape, which can make it overwhelming for both silicon practitioners and academics to adapt to the AI native silicon design era. To add structure to how we navigate this transition, we provide a joint view from academia and industry silicon practitioners of the challenges, opportunities, and considerations we expect will catalyze how the community transitions into an AI native silicon future. In particular, we reemphasize the importance of core silicon design fundamentals in academic training and why they have renewed importance in research and industry practice for AI native silicon design. It is our hope that the views provided here will offer valuable and complementary perspectives to those in academia and industry to interpret, inform, and catalyze the transition to the AI native era. We expect that many similar and overlapping views will emerge, but the precise technical details will differ across stakeholders, so it is valuable for the community to amass a diversity of viewpoints.

\end{abstract}

\maketitle

\fancypagestyle{standardpagestyle}{%
  \fancyhf{}%
  \renewcommand{\headrulewidth}{0pt}%
  \fancyhead[L]{Academia x Industry for Silicon Design}%
  \fancyfoot[C]{\thepage}%
}
\fancypagestyle{firstpagestyle}{%
  \fancyhf{}%
  \renewcommand{\headrulewidth}{0pt}%
  \fancyhead[L]{Academia x Industry for Silicon Design}%
  \fancyfoot[C]{\thepage}%
}
\thispagestyle{firstpagestyle}
\pagestyle{standardpagestyle}

\section{Introduction}
\label{sec:introduction}

Advancements in AI are driving a profound transformation across virtually every major industry, offering unprecedented productivity gains and powerful new design and implementation capabilities. However, fully capturing these gains requires shifting from simple assistive usage to an AI-native paradigm where AI is treated as a first-class consideration in how we operate, solve problems, and design autonomous systems and tools. At the same time, AI's rapid pace of innovation can make it overwhelming to identify durable and general strategies for training AI-native silicon researchers and practitioners. To ground both industry and academia during this transition, we advocate for a renewed focus on silicon design fundamentals and core principles to guide the industry.

Core silicon design first principles, rules, and invariants provide a stable foundation, amidst the rapid pace of change, upon which practitioners and researchers can systematically anchor and build AI-native skills, intuition, and design stacks. Achieving this requires understanding how AI interacts with core silicon design principles and constraints instead of trying to offload fundamental skills and reasoning to AI. Without this foundational mastery, students and practitioners risk succumbing to cognitive offload (i.e., blindly trusting AI results), blind refusal of the technology (i.e., not using AI at all), and implicit knowledge loss over time (i.e., loss of specialized expertise). We are already seeing symptoms of these challenges: (1) in industry practice, inefficient use and overreliance on AI have created significant resource overruns~\cite{forbes_ai_budget, cfodive_ai_cost, fortune_ai_cost}, and (2) in academic training AI is being used to bypass the learning process.

To guard against these pitfalls, students and practitioners should be trained with a new set of AI fluencies and intuitions on top of existing fundamental principles and concepts. This includes understanding how AI interacts with and complements classical silicon design intuition and aptitudes like power, area, performance, and resource usage intuitions. It also includes developing design methodologies on top of existing commercial electronic design automation (EDA) tool flows to get sign-offs but also considering open source and/or surrogate models to build scalable and efficient agentic workflows. Thus, a key theme through this work will be the emphasis on using AI to build complementary solutions on top of core fundamental concepts as opposed to replacing them.

By treating AI fluencies as a complement to existing design approaches, AI can serve as a powerful augmenting capability to both academic and industry practitioners. If designed and used appropriately, agentic AI can be used not only to enhance silicon design productivity but also enable new superintelligence capabilities, scalable technology transfer models, and powerful new emergent design capabilities. These capabilities have the potential to dramatically augment the historical intelligence and expertise capabilities of individual researchers and designers. For instance, superintelligence allows non-experts to harness and deploy expertise for problems where they may not be natively trained, which is a frequent gap in silicon design where roles have historically been less interchangeable than in software. Emergent design capabilities also have the potential to solve historically difficult or engineering-heavy challenges such as lowering a specification to RTL or developing novel optimization techniques.

To prepare the next generation of scientists and engineers for the AI native era, the industry and academia should jointly adapt how we think about training and preparing students for industry practice. This includes adjusting pedagogical practice, rebalancing technical curricula emphasis, and reformulating research and development models. As AI agents absorb routine implementation, debugging, and low-level engineering tasks, technical training should rebalance towards higher-level skills required for agentic thinking and problem solving; these include skills and intuitions like problem decomposition, discerning when (not) to use AI, encoding precise design intent for AI agents, reasoning about randomized processes, and orchestrating and validating AI workflows. Finally, we expect these powerful new capabilities will also enable new opportunities at the frontiers of how we execute, collaborate, and deploy advanced and specialized research expertise across institutions.

\section{Silicon Practice Challenges in the Age of AI}
\label{sec:challenges}

This section highlights some of the key challenges we are seeing for silicon training and practice as we transition to the AI native era.

\subsection{Key Fundamentals for Silicon Design}

Fundamental principles are the core invariants, basic rules, first principles, and theories that remain constant over time. Because these underlying mechanics persist despite advanced AI automation and tools, mastering fundamental theory, best practices, and techniques remains important. With solid fundamentals, AI can then be applied as a complementary layer of capabilities built on top of these immutable rules and best practices instead of a replacement for fundamental aptitudes and skills.
In this work, we refer to fundamentals as both the silicon design principles and skills that more generally transfer across disciplines. More specifically, silicon-specific fundamentals encompass areas such as:

\begin{itemize}
\item \textbf{Computer Architecture:} Defines the overall computer organization structure from instruction set architecture to hardware resource structure, and how programs are executed.
\item \textbf{System-on-Chip (SoC) Architecture \& Interconnects:} Focuses on integrating processors, memory systems, bus protocols (e.g., AXI, AHB), and peripheral IP into a unified system.
\item \textbf{Automation, Simulation, and Modeling:} Teaches the underlying design automation algorithms such as logic synthesis, static timing analysis, and automatic place-and-route, as well as software-based representations of hardware systems with the goal of performance analysis, feature testing, etc.
\item \textbf{Design and Optimization:} Translates functional logic specifications into discrete digital hardware components (logic gates, registers, etc.) and converts them into a precise physical layout of transistors, interconnects, and silicon layers ready for fabrication. Balances processing speed, power consumption, and physical chip size to maximize overall efficiency for specific target applications.
\item \textbf{Verification and Testing:} Rigorously tests and mathematically proves that a hardware design accurately matches its specifications (pre-silicon). Teaches techniques like scan chains, Built-In Self-Test (BIST), and boundary scan to detect physical manufacturing defects in chips (post-silicon).
\item \textbf{Circuit Theory:} Establishes the core mathematical models and physical laws used to analyze and predict the behavior of electrical networks.
\item \textbf{Analog Mixed Signal Circuits:} Designs interfaces that bridge the continuous physical world and discrete digital processing by handling both analog and digital signals on a single chip.
\end{itemize}

In practice, silicon design is highly diversified and deeply specialized so practitioners will typically select one (or a few) of these areas while still learning the basics of the rest of the stack. This creates a unique challenge for silicon design since roles are typically less interchangeable across each specialized field of expertise. This is unlike software design where specialization is less extreme and skill sets are typically more interchangeable across roles (at least at the more junior levels).

In addition to silicon design first principles, we also will refer to general fundamental skills throughout this work which encompass:
\begin{itemize}
\item \textbf{Technical Communication:} Convey complex engineering concepts, trade-offs, and specifications clearly to collaborators through written documentation, diagrams, or presentations.
\item \textbf{Designer Intuition:} Leverage accumulated experience, pattern recognition, and physical principles to quickly narrow down solution spaces and anticipate design bottlenecks before running formal tools.
\item \textbf{Critical Thinking:} Evaluate assumptions, analyze edge cases, and diagnose complex system behaviors logically rather than relying on superficial symptoms or default configurations.
\item \textbf{Precisely Encoding Design Intent:} Translate high-level system requirements and implicit assumptions into clear, unambiguous constraints, assertions, and formal specifications for downstream tools and engineers.
\item \textbf{Systematic Validation:} Methodically plan and execute coverage-driven testing methodologies to systematically stress a design across every reachable state and operational corner.
\item \textbf{Design Justification and Articulation:} Defend engineering choices, trade-offs, and architectural compromises using objective data, metrics, and technically sound reasoning to collaborators.
\end{itemize}

Finally, the next generation of silicon practitioners needs to be equipped with a comprehensive set of AI fluencies to layer AI competencies on top of all of the existing training. One of the key challenges is identifying what AI competencies are valuable especially since AI technologies and leading tools are still fluid. Thus, throughout this work we will attempt to highlight how core foundational principles interact with AI capabilities to connect and contextualize what AI fluencies and concepts are likely important for the AI native era.

\subsection{Implicit Knowledge Loss}

\begin{figure}[t]
  \centering
  \includegraphics[width=\linewidth]{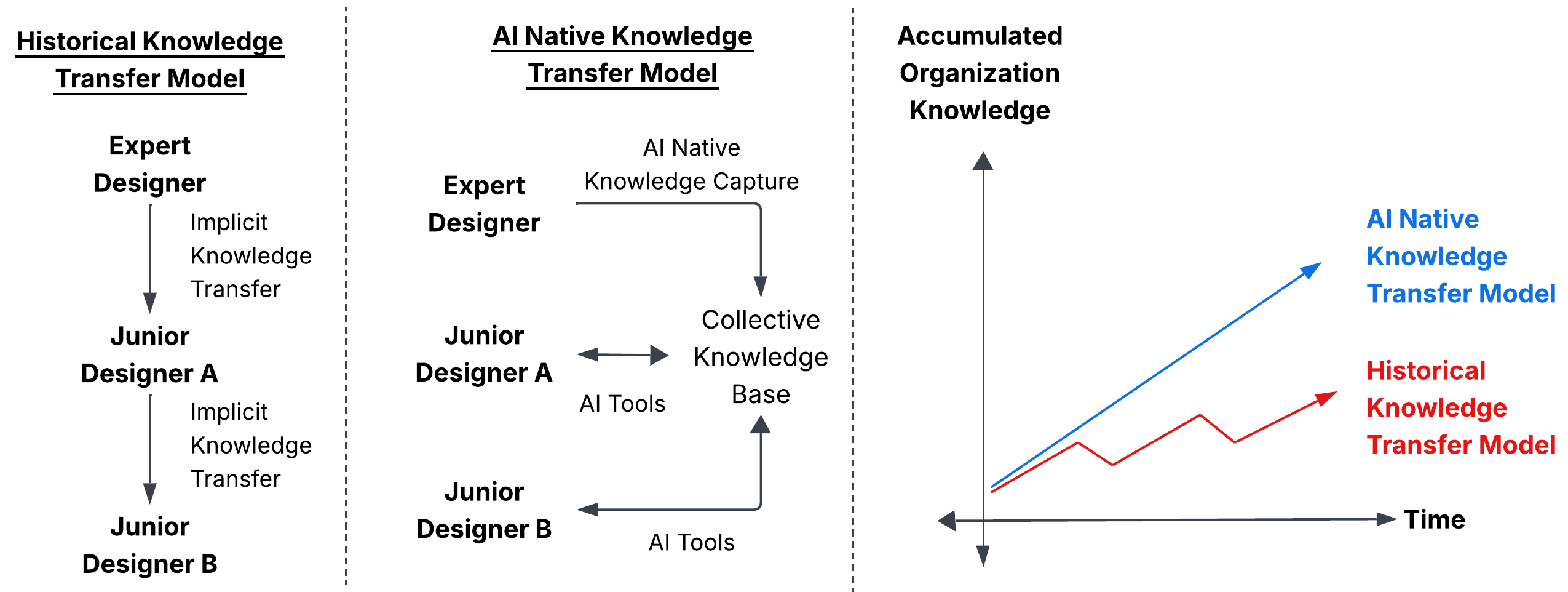}
  \caption{Historical implicit knowledge transfer model risks losing
  knowledge over time while AI native models can reduce
  knowledge loss.}
  \label{fig:implicit-knowledge-loss}
\end{figure}

Silicon design has historically relied on a diverse pool of specialized experts with tribal knowledge across the various design roles (e.g., digital design, verification, DSP, SoC, etc.). As a result, every silicon organization carries a vast body of implicit knowledge in the minds of its engineers such as timing exception rationale, power integrity workarounds for specific technology nodes, and intuition for when design trade-offs are acceptable. This implicit expert knowledge was historically sufficient because the same designers who held it were the ones applying it in practice. However with AI, the shift from executing fundamentals to orchestrating and validating AI-driven workflows exacerbates implicit knowledge loss which risks becoming a bottleneck in AI-native silicon design (\autoref{fig:implicit-knowledge-loss}). In an AI-native workflow, relying on implicit knowledge is fundamentally incompatible with AI workflows since that knowledge is not visible to AI to leverage. This makes it significantly harder to automate and enable the autonomous agentic systems for silicon design.

The impact of knowledge loss also compounds across generations of both engineers and AI models. When an engineer leaves the workforce or changes roles, this implicit knowledge is at risk of disappearing. Traditionally, this knowledge loss process was gradual as other engineers or the team could indirectly absorb fragments through code review, mentorship, and debug sessions. However, in an AI-native workflow the loss is binary: if the knowledge was never encoded in an AI native format, no amount of model improvement can recover the lost expertise. As a result, an AI system may never learn how to automate the task for which the implicit knowledge was lost and instead make decisions that violate unstated assumptions. This risk is further compounded by challenges introduced by AI where a lack of precise requirements can create surfaces for AI agents to hallucinate or make mistakes which can go undetected without the necessary implicit knowledge that was lost.

To guard against these risks, organizations should think about how to capture and encode their silicon fundamentals and tribal knowledge into AI-consumable artifacts. This allows organizations to accumulate and compound knowledge into an AI-compatible format that can be used to improve each successive generation of AI capabilities (models, agents, tools, etc.). Every formalized constraint, specification, and acceptance criterion becomes part of a collective knowledge base that AI agents can reason over and transfer across teams or generations of practitioners. In contrast, organizations that keep knowledge tribal and distributed across individual engineers’ expertise will find that expertise increasingly difficult to scale and transfer, and harder to protect against knowledge loss. In addition, since gains in AI capabilities compound over time, the gap between these two operational modes will widen with every model generation.

We also expect this challenge to scale beyond any single organization to the industry. Without preserving implicit knowledge across industry sectors, there is a high risk that if a company disappears tomorrow, a large amount of implicit knowledge will also permanently disappear. Guarding against this industry-scale knowledge loss will similarly require rethinking how we capture, communicate, and preserve this knowledge. To do this reliably, we should standardized, machine-readable formats to express specifications, constraints, assumptions, requirements, and acceptance criteria to enable shared interchange similar to existing standardized communication protocols.

\subsection{Cognitive Offload}

\begin{figure}[t]
  \centering
  \includegraphics[width=\linewidth]{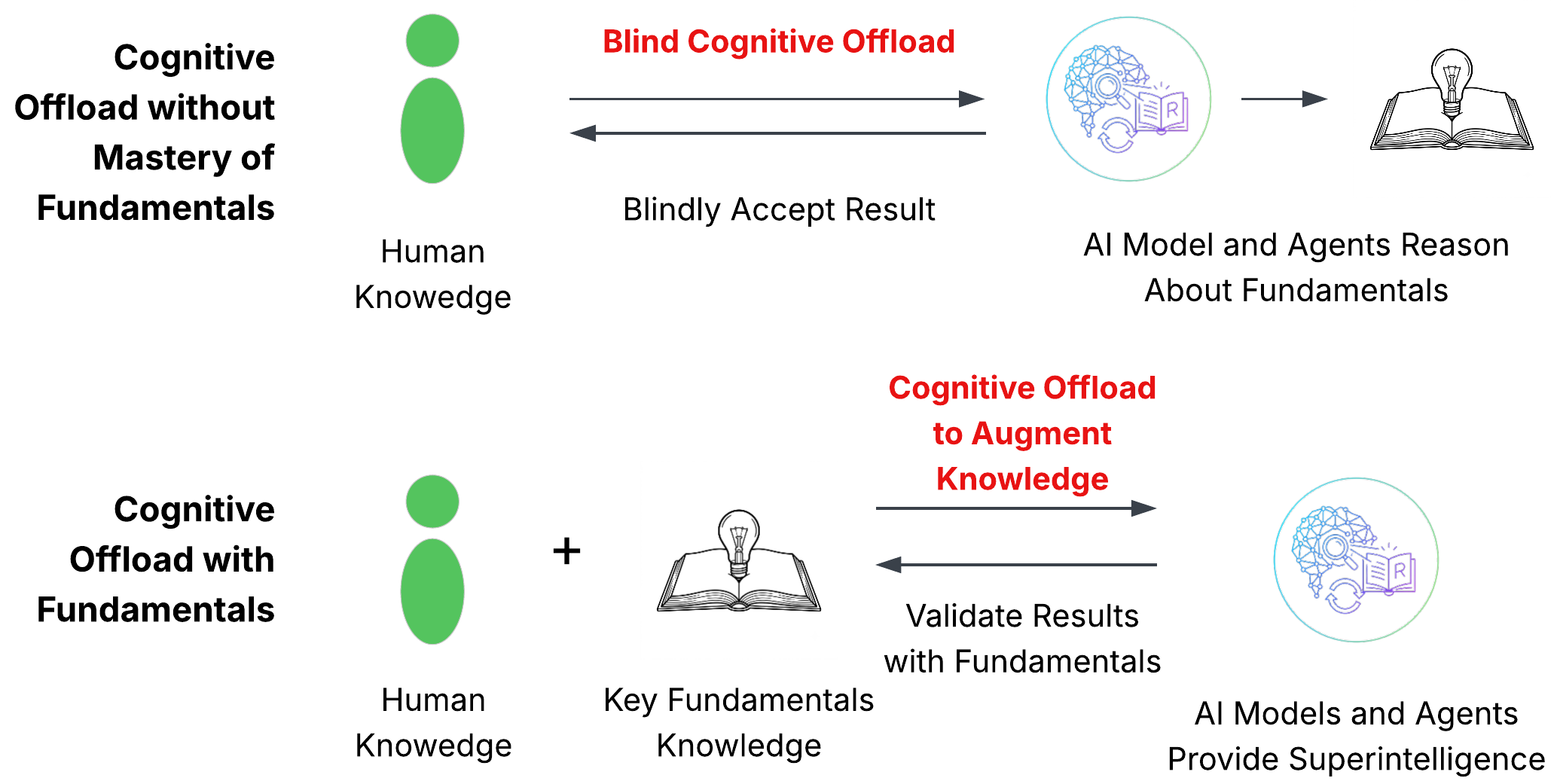}
  \caption{Cognitive offload to bypass mastery of fundamentals makes it
  difficult to validate results. Cognitive offload, if used correctly, can enable superintelligence to augment designer expertise.}
  \label{fig:cognitive-offload}
\end{figure}

AI tools create new risks that fundamental concepts and expertise are never fully mastered due to overreliance on and cognitive offload to AI. Cognitive offload is defined as the practice of delegating tasks externally (to the AI) to avoid having to conduct the work explicitly. If applied correctly, cognitive offload can be a powerful way to augment designer productivity (e.g., superintelligence); however, excessive cognitive offload carries the risk of eroding a designer's ability to validate, think critically, develop efficient solutions, and capture potential bugs (\autoref{fig:cognitive-offload}). This is problematic when implementing the precise checks and invariants required to ensure airtight correctness necessary for silicon design. Software engineering faces a similar challenge but the issue is far more important in silicon design. In software, bugs can be patched post-deployment due to the significantly lower cost of remediation, whereas fixing bugs in silicon is exceptionally resource intensive.

One way cognitive offload manifests in industry practice is the blind application of AI to formulate solutions, offload tasks, and bypass fundamental best practices and checks. For instance, we are seeing many instances of both AI-generated designs and tests which violate the historical software and silicon development best practice of keeping these processes separate. We also see that tasks which have existing deterministic and known solutions are being blindly offloaded to effectively reinvent solutions with AI which are more resource intensive and less reliable. Finally, we also see that with AI students and designers sometimes cannot explain why the solution the AI generated works; this undermines trust in the generated collateral and typically does not meet the airtight validation guarantees needed in silicon design.

In academic settings, AI tools make it easy to bypass the productive friction that builds mastery. Working through a problem set, getting it wrong, and reasoning about why is how core concepts and foundational principles get internalized, yet AI can now produce a correct-looking answer before that process ever starts. Left unaddressed, this puts at risk the very foundation upon which advanced research and development capabilities are built. This is particularly problematic for AI native silicon design practice which will require understanding of all of the previous skills, foundational principles, and technical expertise of silicon design in addition to new AI techniques and tools.

To guard against the pitfalls of cognitive offload, silicon designers will need to have the same deep engagement with fundamentals that AI tempts students and engineers to bypass. As a practitioner, you cannot judge whether AI made a good design decision if you never struggled with making that decision firsthand and understanding whether the solution was correct yourself. It is also difficult to fully specify a task and prompt the AI with all of its technical parameters if you never executed or understood the task yourself.

\subsection{Shifts in Silicon Training: Bottom-Up vs. Top-Down}

\begin{figure}[t]
  \centering
  \includegraphics[width=\linewidth]{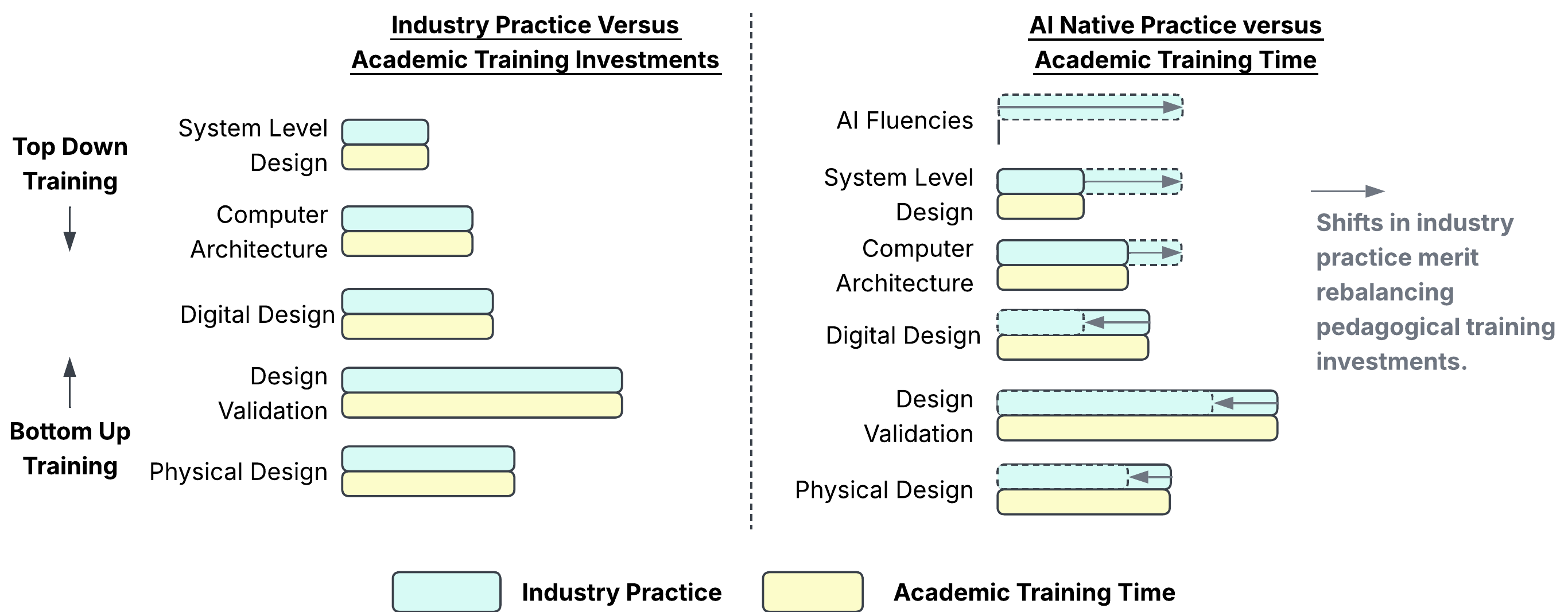}
  \caption{Before agentic AI, academic training roughly aligned with
  industry practice workflow needs. Shifts in industry practice to AI native silicon design may require rebalancing
  training investments\protect\footnotemark.}
  \label{fig:top-down-bottom-up}
\end{figure}

Silicon design has typically been structured in both bottom-up and top-down views (\autoref{fig:top-down-bottom-up}). Bottom-up coursework such as device physics, circuit theory, digital logic, and RTL design builds foundational intuition about how physical constraints propagate upwards through abstraction layers. Top-down coursework, like system architecture, specification-driven design, and project-based capstones, develops the ability to reason about intent, constraints, and trade-offs at the system level. Over years of practice and refinement, curricula have been attuned to roughly match the demands of industry practice.

However, as AI agents accelerate the design process, where practitioners spend their time in the design stack and workflow will shift (likely unevenly). Part of this shift will be towards higher-level top-down formulation of system-level design intent, defining constraints and acceptance criteria, orchestrating agent workflows, and validating design results. Bottom-up understanding will remain important for foundational basics and lower-level technical areas where AI is not yet effective. To adapt, curricula will need to rebalance and adapt how much focus is spent on bottom-up training versus top-down higher-level reasoning and orchestration required to operate AI native workflows.

\footnotetext{How much each area of training is adjusted is still an open
question. The magnitude of shifts here is for illustrative purposes.}

Defining what constitutes ``minimally sufficient'' mastery in each technical area is a non-trivial challenge; designers will need enough knowledge across the design stack to understand why constraints exist, to recognize when AI-generated outputs violate physical reality, and to formulate meaningful evaluation criteria that can be captured in deterministic software. They need to have struggled with timing closure, power integrity, and functional verification at least once to develop the judgment required to assess whether an AI agent's solution is trustworthy. The precise boundary will evolve as AI capabilities advance, but at a high level it is likely bottom-up mastery will need to be calibrated to what is necessary for effective top-down reasoning and validation, not to what was historically necessary for manual implementation.

A more practical challenge is the accessibility and visibility to university settings in how top-down silicon design tasks are practiced in industry. Top-down silicon design reasoning is practiced almost exclusively in industry on highly complex proprietary product specifications under business operating constraints. Production chip programs can take multiple years and architectural decisions constrain every downstream design decision which is difficult to capture in a single academic course. Addressing this gap will require tighter collaboration (e.g., internships, training programs, etc.) between industry and academia to provide representative exposure to top-down design reasoning in academic settings to rebalance curricula.

The risk of not rebalancing training is that graduates may be technically proficient at manual implementation but unprepared for the orchestration and validation roles in professional practice. At the other extreme, it risks training designers who can specify what they want but cannot recognize when the result is wrong. Neither extreme is desirable so we need to be open to intentionally balance and continuously recalibrate training concentration as AI proficiencies in industry practice evolve.

\subsection{The Structural Data Scarcity Problem}

\begin{figure}[t]
  \centering
  \includegraphics[width=\linewidth]{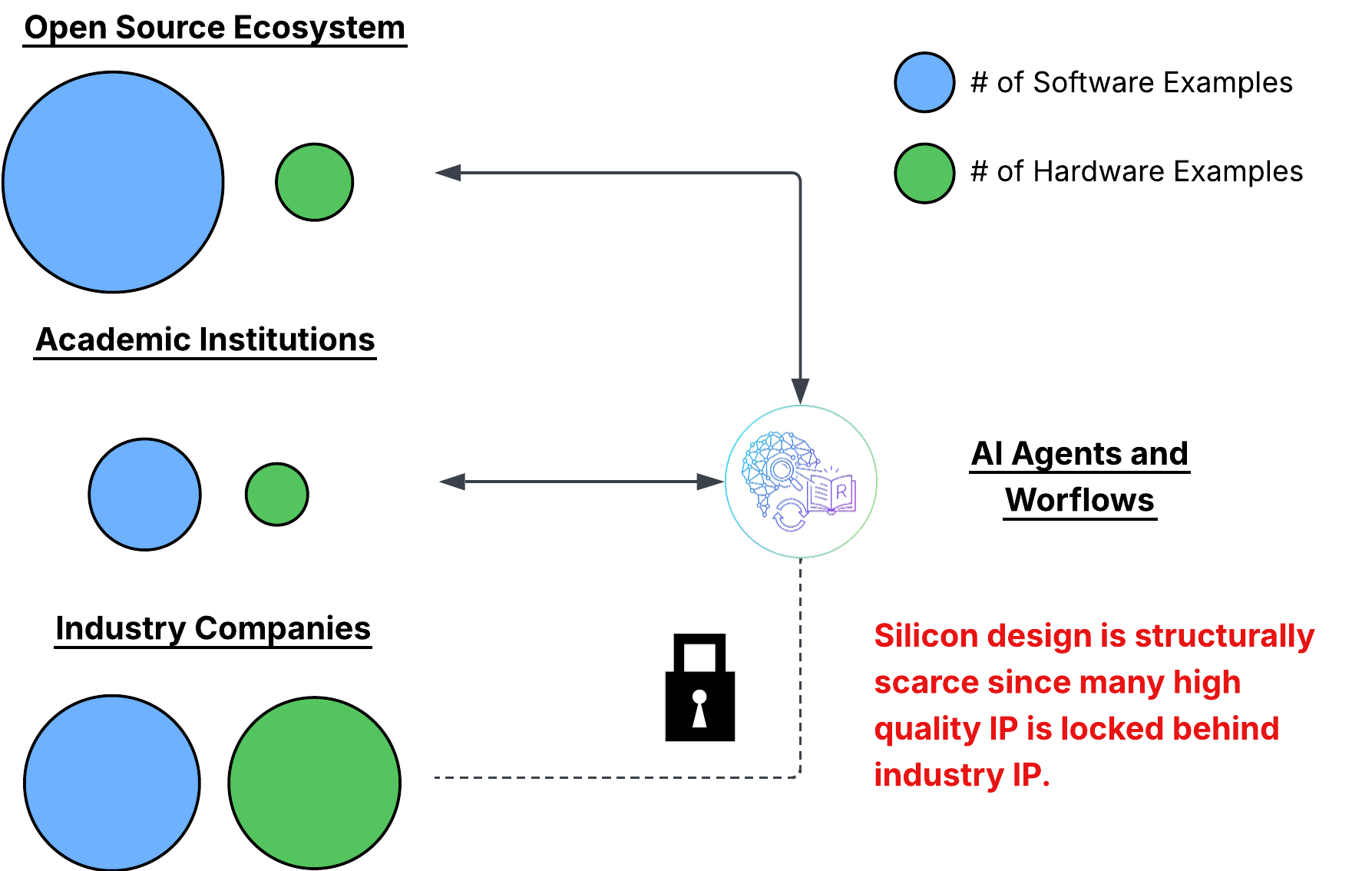}
  \caption{Compared to software, the availability of silicon designs is
  structurally scarce, which limits traditional AI scaling laws that rely on
  vast amounts of data to improve model performance.}
  \label{fig:structural-data-scarcity}
\end{figure}

A key difference between software and silicon for AI development is the availability of training data (\autoref{fig:structural-data-scarcity}). In software, open-source repositories provide billions of lines of code, bug reports, test suites, and execution traces that can be scraped, labeled, and harnessed to train increasingly advanced LLMs. In contrast, high-quality silicon data is comparatively and structurally scarce since it is locked behind NDAs, coupled to specific technology nodes, fragmented across vendor-specific formats, and rarely shared outside industry repositories. This initially limits the traditional AI scaling laws which rely on the availability of additional data to improve model performance since the data simply does not exist in sufficient quantity or diversity (at least not yet).

The structural data scarcity problem is also deeper than simply available RTL IP designs. Each stage in the chip design flow also requires or produces intermediary collateral such as specification documents, EDA tool logs, bug reports, and timing analyses which engineers use to drive educated design decisions. This design process and collateral are rich with intermediary signals and data that would need to be harnessed to formulate intelligence to drive silicon design agents. The problem is further exacerbated since many of the key gates and checkpoints require implicit knowledge for tasks like sign-off analysis and judging whether results are sufficient such as whether timing margins are reasonable, if it is acceptable to waive DRC violations, or whether verification coverage adequately covers corner cases.

Converting these human judgment calls into deterministic, machine-evaluable checks is already a substantial engineering challenge. Even where organizations succeed in formalizing these checks into deterministic evaluation criteria, the resulting artifacts can be internal and often require custom tool flows/methodologies to evaluate which are also guarded by the same NDAs and IP protections. This makes the structural scarcity problem multilayered since not only is the raw design and intermediary data scarce and likely internal, but the evaluation infrastructure needed to grade any AI agents is equally scarce and likely idiosyncratic. This makes enabling the flywheel to improve AI models for silicon structurally more challenging; as a result, any viable AI strategy for silicon will have to account for and resolve these challenges to fully enable transformational silicon design capabilities.

\subsection{The Rate-of-Change Problem for Curriculum}

As AI rapidly alters the technology landscape, the historical challenge of keeping academic curricula up to date has been drastically magnified by the sheer velocity of AI innovation and pace of change. For academic training this poses the question: “How do we adapt academic training when the pace of change outstrips traditional curriculum cycles?” This dilemma is especially acute as the frontier of knowledge rapidly expands due to AI and advanced techniques more quickly solidify as necessary foundational skills. To bridge this gap between industry practice and classroom training, we will need to revisit how we teach, what we teach, and how we formulate what we teach.

At the core of this pedagogical transformation is a shift in teaching methodology. The rapid evolution of AI tools introduces significant complexity ranging from navigating fast-changing software versions to managing the non-deterministic and sometimes highly personalized outputs of AI models. This can clash with how we teach the basics which is often done in deterministic, precise, and predictable environments. To bridge the gap, curricula will need to build upon existing structured course progression to absorb AI topics: establishing core digital design fundamentals, advancing to non-deterministic AI-assisted workflows, progressing to fully autonomous agentic design, and ultimately advanced, state-of-the-art topics.

To accommodate this expanded scope without lengthening degree programs, educators will likely need to carefully consider leveraging AI itself to optimize the learning process. By integrating AI-assisted instruction under close faculty guidance, students can digest and master foundational concepts in a compressed timeframe. The hours saved through this accelerated core instruction can then be reinvested into hands-on application or additional advanced course material (such as AI training). AI can also be used to update curriculum materials but human oversight will remain vital to ensure that the fundamental facts and principles remain correct since these will ultimately be used to drive the remainder of the students' careers. How the specific mechanics of these updated teaching approaches manifest for silicon design is an open question but the community will need to work through them as we adapt to the AI native era.

\subsection{AI Native Technology Stack Evolution}

The semiconductor industry over decades has evolved as innovation and research were gradually encapsulated, democratized, and commoditized as automated tools to enable new innovations to be layered on top of them.
For instance, EDA tools encapsulated the advanced expertise for tasks like logic synthesis, placement and route, and manufacturing DRC checks which enabled engineers to leverage built-in expertise.
Similarly, accompanying IPs like PCIe, USB, and CPU processors encapsulated the deep microarchitecture and optimizations which allowed designers to leverage high-performance IP without having to build it themselves to support complex software stacks and systems.
This encapsulation enabled the industry to democratize expertise so that anyone could contribute to silicon design by leveraging, composing, and reusing the technology stack that would otherwise be difficult for any individual engineer or institution to fully assimilate.

The AI native shift provides a similar opportunity to hierarchically build agentic AI on top of the existing silicon design technology stack without having to start from scratch.
Currently, it is still an open question as to what the final form of these new AI native encapsulated technologies will look like since the supporting AI models, tools, frameworks, concepts, and infrastructure are rapidly changing.
However, it will be important to ensure that any AI native silicon tooling or automations are composed on top of existing known best practices and technologies in a complementary way to leverage knowledge embodied in existing automation.
To enable this, the industry should establish the frameworks, standards, and methodology to bring these new AI native silicon technologies to a similar level of stability and reproducibility necessary to commoditize and democratize the agentic AI contributions to the technology stack.

\section{Changes to Silicon Designer Roles}
\label{sec:roles}

This section highlights some of the projected shifts in silicon design roles in the AI native era.

\subsection{Gradual Commoditization of Code Generation for Silicon}

\begin{figure*}[t]
  \centering
  \begin{minipage}{\textwidth}
    \centering
    \begin{subfigure}{0.48\textwidth}
      \centering
      \includegraphics[width=\linewidth]{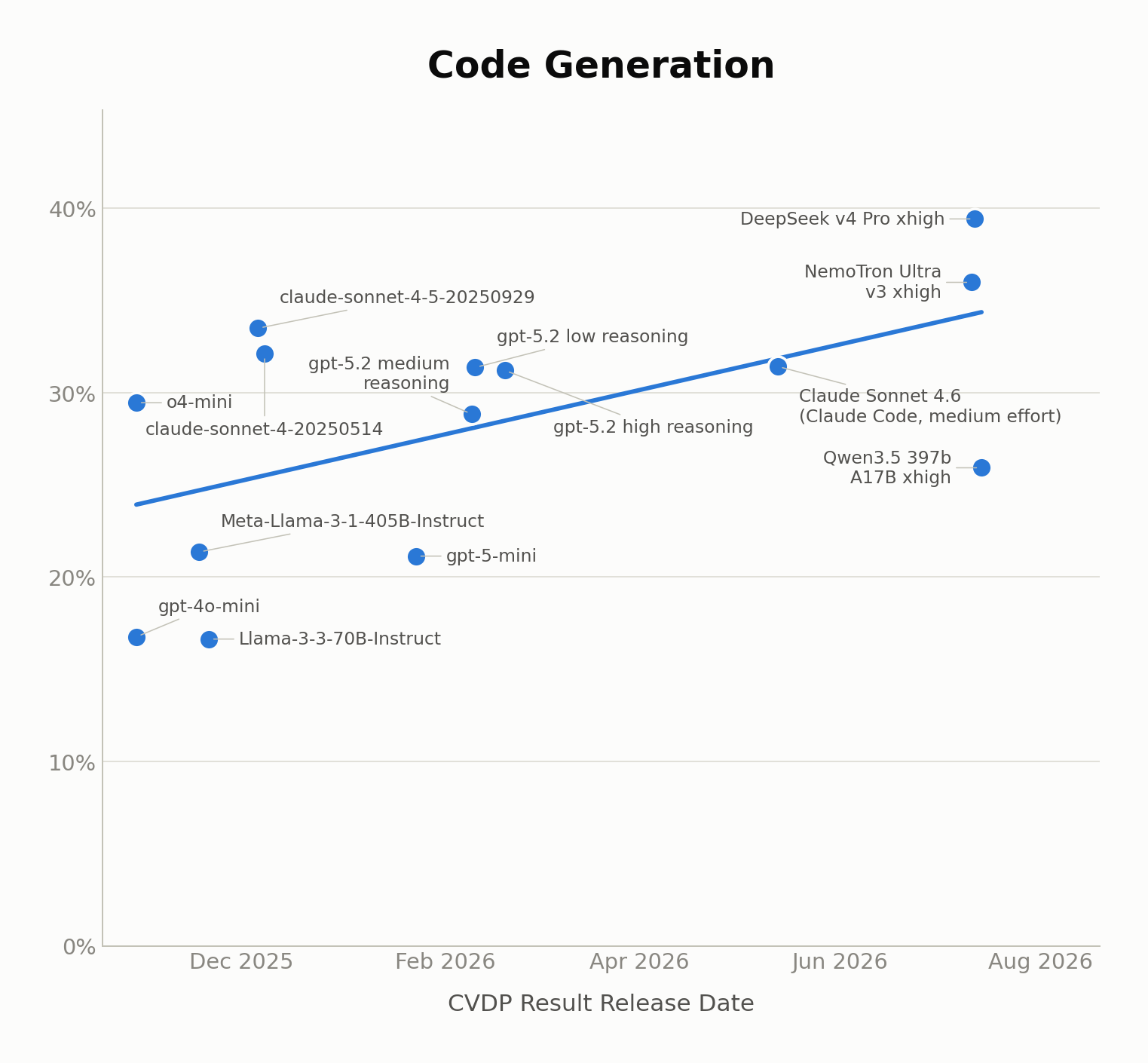}
    \end{subfigure}
    \hfill
    \begin{subfigure}{0.48\textwidth}
      \centering
      \includegraphics[width=\linewidth]{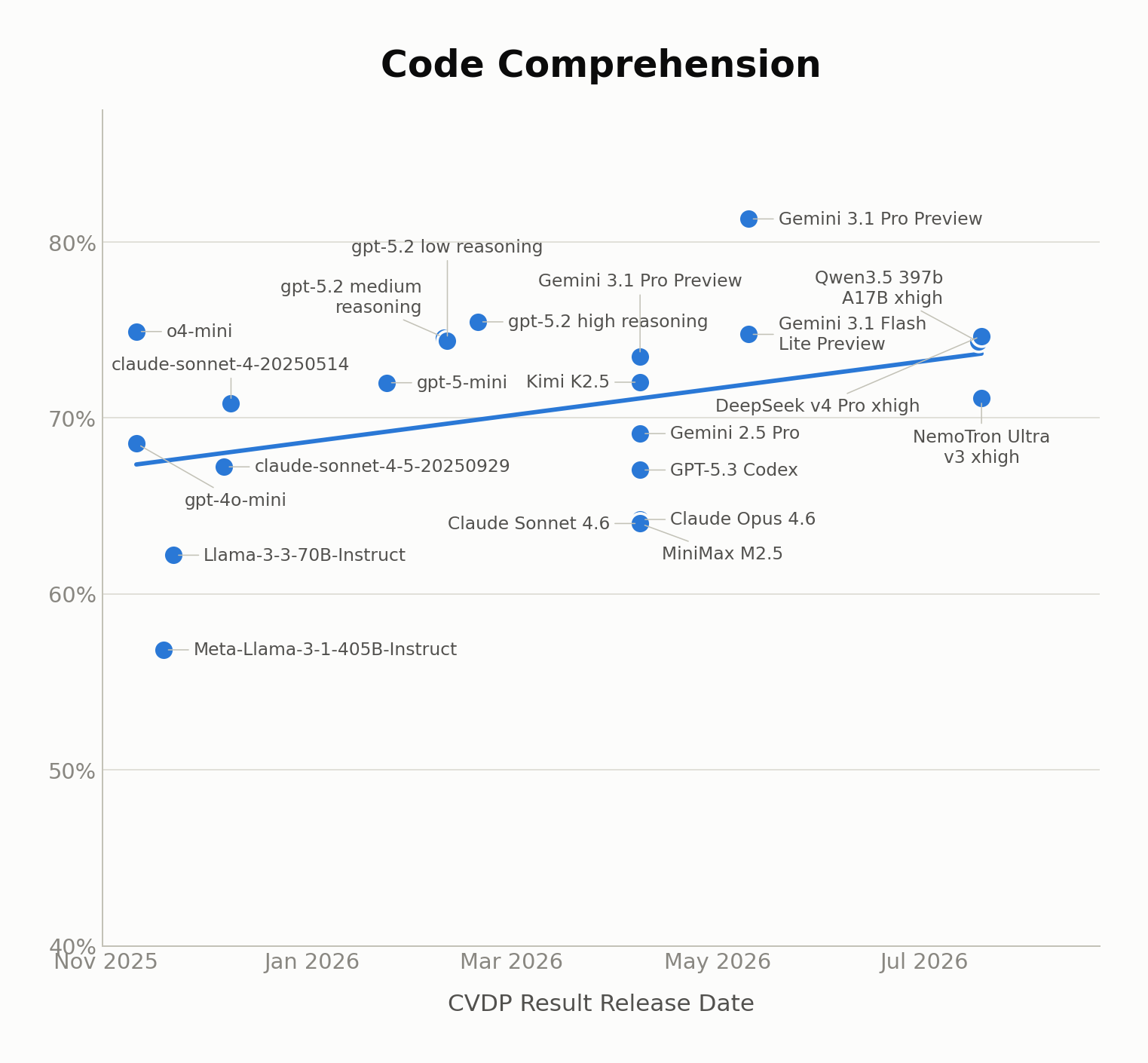}
    \end{subfigure}
    \caption{CVDP performance (\% Resolved) of frontier models over time for
    RTL code generation tasks and code comprehension tasks as of
    August 2026~\cite{cvdp_leaderboard}. Higher is
    better.\protect\footnotemark}
    \label{fig:cvdp-performance}
    \renewcommand{\thempfootnote}{\arabic{footnote}}
    \footnotetext{Includes data that was submitted but neither run nor
    verified directly by Si2.}
  \end{minipage}
\end{figure*}

Advanced LLMs excel at code generation and seamlessly converting between programming language representations. For instance, AI coding agents are now generally good at converting from natural language specifications to most commonly used programming languages like Python and C++. In silicon design, frontier LLMs currently leave some capabilities to be desired due to the scarcity of silicon training data. However, we expect that this will change and we are starting to see improvements in the efficacy of RTL code generation by LLMs. \autoref{fig:cvdp-performance} shows the results of several generations of frontier models on the CVDP~\cite{cvdp} code generation and code comprehension benchmark.

At a high level, we expect frontier models to gradually commoditize code generation for RTL based on the data and trends as models become increasingly capable. As a result, we expect that the role of a silicon designer will also gradually shift towards higher-level orchestration tasks similar to how software code commoditization has changed software engineering roles. In the meantime, we can bridge the RTL code generation shortcomings with transfer learning, in-context learning, and other techniques which show reasonable results for various RTL tasks~\cite{autochip, verilogeval_revisited, verilogcoder}. In the long term, as LLM performance for RTL code generation improves, the industry should revisit the role and value proposition of different silicon design languages and hardware DSLs. For instance, if specification-to-RTL is eventually commoditized, then the productivity value propositions for domain-specific, higher-level hardware description, or HLS languages should be rethought and reassessed.

\subsection{Shifting Silicon Workflow Bottlenecks}
\label{subsec:workflow-bottlenecks}

AI has dramatically accelerated the speed at which certain tasks such as code generation, test implementation, and debugging are executed. As a result, workflow bottlenecks are gradually shifting to other areas like planning, critical thinking, communication, review, and validation, making these types of tasks increasingly important to master. To illustrate the rough magnitude of this shift, we refer to a 2019 study at Microsoft (prior to disruptions due to COVID and AI) where researchers analyzed where 5,928 software developers spent most of their work time.

\begin{table*}[t]
  \centering
  \caption{Workflow breakdown for N=5,928 software engineers~\cite{meyer2019} and projected shift in bottlenecks with 2, 5, and 10x AI productivity gains for implementation tasks. Aspects like technical communication and soft skills become increasingly important.}
  \label{tab:workflow-bottleneck}
  \begin{tabular}{@{}p{3.4cm}lcccc@{}}
    \toprule
    & \textbf{Activity}
    & \textbf{\shortstack{Workflow\\Share (\%)}}
    & \textbf{\shortstack{Projected Share\\(\%) @ $2\times$}}
    & \textbf{\shortstack{Projected Share\\(\%) @ $5\times$}}
    & \textbf{\shortstack{Projected Share\\(\%) @ $10\times$}} \\
    \midrule
    \multirow{5}{3.4cm}{Implementation Tasks Accelerated by AI}
      & Coding          & 15 & 9.6 & 4.6 & 2.5 \\
      & Bugfixing       & 14 & 9.0 & 4.3 & 2.3 \\
      & Testing         &  8 & 5.1 & 2.5 & 1.3 \\
      & Reviewing Code  &  5 & 3.2 & 1.5 & 0.8 \\
      & Documentation   &  2 & 1.3 & 0.6 & 0.3 \\
    \midrule
    \multirow{10}{3.4cm}{Other Workflow Tasks (i.e., everything else)}
      & Specification        &  4 &  5.1 &  6.2 &  6.6 \\
      & Meetings             & 15 & 19.2 & 23.1 & 24.8 \\
      & Email                & 10 & 12.8 & 15.4 & 16.6 \\
      & Interruptions        &  4 &  5.1 &  6.2 &  6.6 \\
      & Helping              &  5 &  6.4 &  7.7 &  8.3 \\
      & Networking           &  2 &  2.6 &  3.1 &  3.3 \\
      & Learning             &  3 &  3.8 &  4.6 &  5.0 \\
      & Administrative Tasks &  2 &  2.6 &  3.1 &  3.3 \\
      & Breaks               &  8 & 10.3 & 12.3 & 13.2 \\
      & Various              &  3 &  3.8 &  4.6 &  5.0 \\
    \bottomrule
  \end{tabular}
\end{table*}

The results of this study (\autoref{tab:workflow-bottleneck}) showed that development-heavy implementation tasks like coding, bugfixing, and testing, as well as tasks that can be AI-assisted like documentation and code review typically comprised around 44\% of a software developer’s time~\cite{meyer2019}. This means that even if agentic AI speeds up implementation by one order of magnitude (say 10x), the remaining workflow bottlenecks will limit the overall productivity gains to 1.66x. While analogous data for silicon design does not exist, we expect that at a high level the trends will be similar. More broadly, this trend means the industry and academia should reemphasize training for the fundamental “soft” skills required to accelerate the remaining workflow tasks such as technical communication, articulation, and collaboration skills.

\subsection{Formulating and Harnessing Superintelligence}

Silicon design requires a wide breadth of specialized experts spread across various roles to come together to build a chip. These roles have historically not been directly interchangeable due to the narrow and deep specialization required for each silicon design task; for instance, an RTL design verification engineer cannot simply take the role of a firmware engineer or power architect without some effort. This makes silicon design roles and resources less interchangeable than in software engineering, where the concentration and variance of roles are narrower.

\begin{figure}[t]
  \centering
  \includegraphics[width=\linewidth]{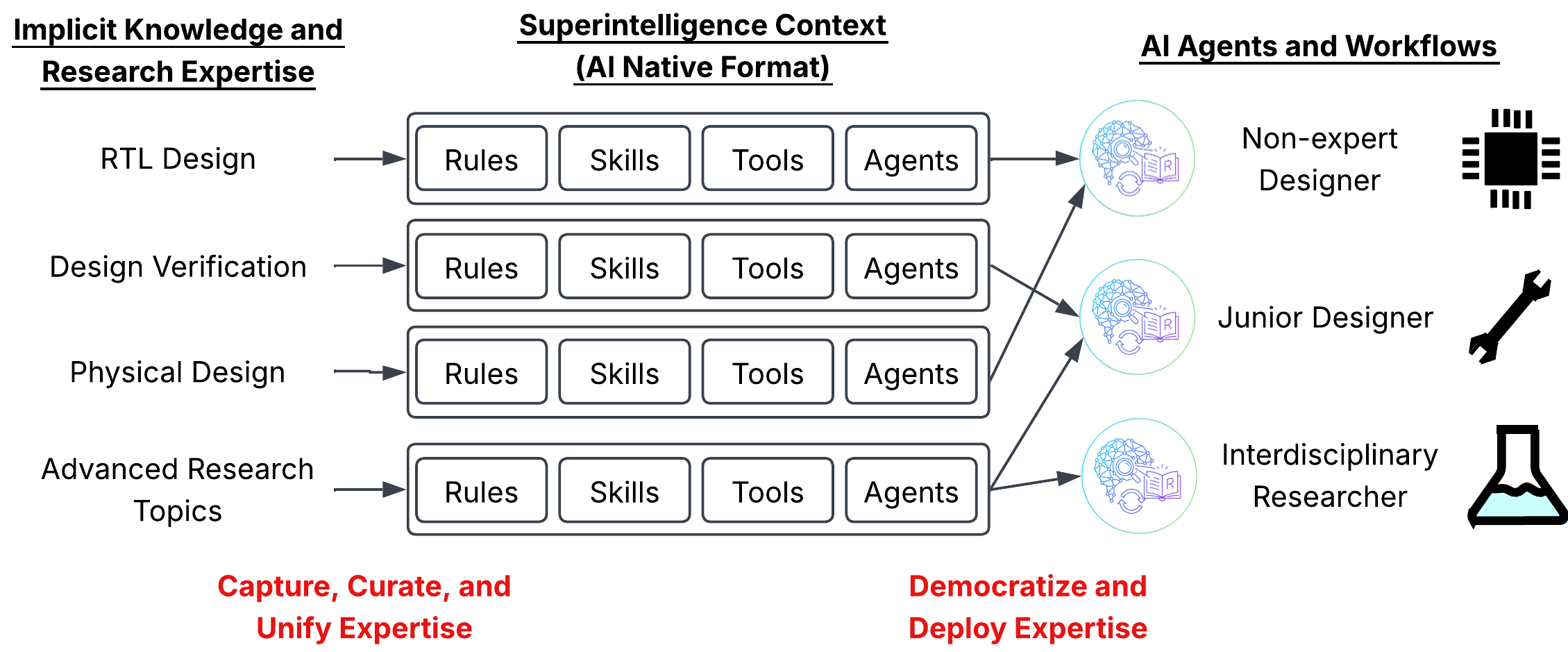}
  \caption{Expertise captured as superintelligence in an AI native format can be deployed at
  scale to augment designer
  expertise and guard against implicit knowledge loss.}
  \label{fig:superintelligence}
\end{figure}

To address this and guard against knowledge loss, silicon practitioners should understand how to formulate and harness superintelligence (\autoref{fig:superintelligence}). Superintelligence is the concept of augmenting an individual designer's intelligence and capabilities with AI tools and context in which the designer may not be a native expert. For instance, by harnessing superintelligence about vector architectures, a non-expert may be able to apply vector code optimizations without knowing the details of the particular architecture or ISA intrinsics. This sort of superintelligence is made possible because agentic AI provides a scalable mechanism to capture and absorb specialized expertise for any design workflow via context, skills, tools, agents, etc. The additive nature of these mechanisms enables a form of collective superintelligence context that is greater than the sum of the individual expertise pools.

This superintelligence can then be deployed and democratized across non-experts to augment the individual designer capabilities. In particular, this allows silicon designers to become more elastic in their roles and better support the cyclical development behavior throughout a chip program’s lifecycle. The full capabilities and limitations of this type of collective superintelligence are not well known; however, it will still be important for silicon practitioners to understand what kinds of capabilities this technology confers as well as how to construct, deploy, and apply these capabilities.

\subsection{Leveraging Formal Correctness as Reward for Agents}

Despite the structural data scarcity problem, silicon design has a key structural advantage in that many of the tasks have existing mathematically precise correctness criteria that can be leveraged. For instance, areas like functional verification already employ equivalence checking and formal verification techniques which mathematically ensure transformations preserve semantics (albeit with potentially long setup run times). Static timing analysis tools yield highly precise deterministic pass/fail checks against timing constraints, and design rule checks enforce physical manufacturability with binary precision. Each of these checks is empirical, deterministic, and precise, and provides ground-truth reward signals where correctness is backed by mathematics (i.e., no subjective human annotation). Thus, designers should transfer and reapply these formalized checks to agentic silicon workflows to leverage existing and formalized reward functions.

If properly applied, the evaluation and graders required to guide agentic AI workflows and design can still be developed using these formalized checks despite the scarcity of labeled data that would otherwise be needed. For instance, low-level design tasks like lint, syntax validation, and basic constraint checking are typically fast and can provide rapid reward feedback to agentic workflows. On the other extreme, full sign-off flows like timing closure, DRC, and LVS provide the formal checks to gate the design for production. Since each signal is a deterministic grader, it enables a way to reward AI agents and provides feedback on the quality and correctness of any AI-generated collateral without requiring pre-existing training corpora. As a result, designers should be able to identify and leverage existing deterministic evaluation signals throughout their workflows to guide AI agent design in lieu of labeled data.

\subsection{Validating and Explaining AI-Generated Collateral}
\label{subsec:validating-ai-collateral}

As increasing amounts of collateral are generated by AI workflows, designers should shift more time to being able to interpret, trust, and explain the results. If we provide purely functional-correctness or PPA-based reward signals to AI agents, there is no mechanism constraining the agent to follow human interpretable patterns or conventional design methodologies. On the other hand, there are also benefits to loosening constraints since the agent is more free to discover circuit topologies, microarchitectural organizations, or physical implementations that no human engineer would have conceived.

These considerations raise fundamental questions about validation and trust that the silicon community has yet to confront at scale. When a human implements a design, reviewers can trace the design rationale, map it against known topologies, and draw on pattern recognition built from years of experience. When an AI agent produces a functionally correct design that follows no recognizable pattern, these conventional validation mechanisms break down and make it harder to trust the design. As a result, designers in an AI native era should become more familiar with how to: (1) verify design behavior and formulate comprehensive checks to catch novel failures, and (2) provide explainable summaries (of potentially novel patterns) to verify and justify AI-generated design decisions.

\section{Building AI Native Silicon Intuition}
\label{sec:intuition}

In the AI native future, silicon design practitioners and researchers will need to build AI intuition across both classical areas and new AI skills.

\subsection{AI x Classical Silicon Design Intuitions}
\label{subsec:classical-intuitions}

Silicon designers over years of academic training and practical experience develop intuition as to how to optimize their workflows and design metrics such as power, performance, area, and resource use which enables almost reflexive design decision making. In the AI native era, silicon designers will also need to develop intuition as to how AI interacts and augments with these classical silicon design intuitions. To make this concrete, it is important to understand how we built these classical intuitions to draw parallels to how they interact with and compose with AI (\autoref{tab:intuitions}).

\begin{table*}[t]
  \centering
  \caption{Classical silicon design intuitions and examples of their AI native
  counterparts. Practitioners need to understand how AI interacts with existing intuitions to make AI application to silicon design reflexive.}
  \label{tab:intuitions}
  \small
  \begin{tabular}{@{}>{\raggedright\arraybackslash}p{0.14\textwidth}
                    >{\raggedright\arraybackslash}p{0.40\textwidth}
                    >{\raggedright\arraybackslash}p{0.40\textwidth}@{}}
    \toprule
    \textbf{Classical Intuition Type}
    & \textbf{Silicon Design Intuition Example}
    & \textbf{AI Native Silicon Intuition Example} \\
    \midrule
    Workflow
    & An EDA tool run for my design takes 8 hours so I will structure my
      workflow to run the tools overnight so I can inspect the results in the
      morning.
    & This silicon design task is relatively easy for the AI so I'll run it
      with lower effort instead of high or max effort to reduce run time. \\
    \addlinespace
    Power
    & If I reduce the active duty cycle of the workload, the silicon will
      consume less power so I want to look for duty cycling opportunities.
    & I want the AI to optimize the design power autonomously. Here are the
      techniques the AI can try and how to evaluate autonomous
      optimizations. \\
    \addlinespace
    Performance
    & Inserting additional pipeline registers can improve the clock speed and
      performance of the design so I will look for opportunities along the
      critical paths.
    & The AI agent can iteratively analyze timing slack across thousands of
      paths simultaneously, so I will prompt it to insert retiming registers
      and re-balance pipeline stages along critical timing paths
      autonomously. \\
    \addlinespace
    Area
    & Sharing hardware logic across different operations or reusing memory
      blocks reduces total die size, so look for opportunities to
      time-multiplex resource-heavy components.
    & Since the AI agent can explore and find non-obvious optimizations I'll
      have it autonomously explore the vast microarchitectural design spaces in
      parallel for non-obvious area tradeoffs for floorplanning and
      logic-sharing. \\
    \addlinespace
    Resource Usage
    & I need to make sure that
      we use EDA tools sparingly and only for tasks that require sign-off quality
      results.
    & Tokens are resource intensive so I need to engineer my prompts to express the
      silicon design specifications as efficiently as possible. \\
    \addlinespace
    Design Specification and Validation
    & To fully specify an IP block design, the design specification needs to
      document the ports, interface bit width, internal functionality, etc.
    & To fully specify a task for the AI, I need to provide these technical
      details and specifications in the prompt to ensure the AI does not
      hallucinate or make potentially incorrect assumptions. \\
    \bottomrule
  \end{tabular}
\end{table*}

\smallskip \noindent \textbf{Workflow Intuition.} A distinguishing aspect of silicon design (compared to software) is the reprecussions of failure and long-running, license-restricted EDA/CAD tool flows. This has made historical silicon design workflows bottlenecked by resource-intensive tool flow runs which are typically used sparingly to optimize run time. Through experience, designers build intuition on roughly how long a synthesis or place-and-route workflow will take, and which tasks and tools run quickly. This run time intuition allows us to structure and parallelize silicon design workflows to minimize disruption and maximize workflow throughput. For instance, we may batch run multi-hour EDA/CAD flows overnight and look at the results the next morning as opposed to launching it in the middle of the day and letting the results idle after hours.

AI now provides both new opportunities and challenges since on the one hand AI can now orchestrate these flows automatically but on the other hand potentially exacerbate the run times. Automated workflow orchestration can be valuable to continuously invoke and monitor long-running jobs but since AI is error-prone it also risks increasing the number of failed EDA tool flow invocations. The run time of AI agents can also be unpredictable (e.g., agentic loops) which introduces additional workflow optimization challenges. To build this intuition and experience, practitioners will need to intentionally experiment and monitor how AI tools behave and interact with EDA tools for various design tasks similar to how we learned to master working with long-running EDA/CAD tools.

\smallskip \noindent \textbf{Power, Performance, and Area Intuition.} Historically as silicon designers we have strived to balance and optimize power, performance, and area (PPA). The intuition of how different design decisions and techniques correlate to PPA improvements over years of training is now almost reflexive. For instance, if designs do not meet clock frequency targets, we look at our toolkit of optimization techniques such as adding retiming or pipelining registers; if we need to optimize power, we look to techniques like clock gating and DVFS. If we need to minimize area, we try to reuse modules and circuitry to reduce resource requirements. These techniques and approaches are all well-established and are the culmination of years of training and experience.

With advanced AI, silicon designers should update their toolkit to absorb how to apply AI to complement existing intuition and techniques for PPA optimization tasks. This requires strong fundamentals training since designers need to understand the basic mechanics, utility, and impact of existing techniques on PPA before AI can be applied to further achieve additional gains. For instance, recent advancements in AI tools allow users to encode recipes and techniques in skill files that the AI can leverage to orchestrate and execute tasks. However, for these skills to be maximally effective the designers should both express the existing technical intuition as to (1) when the skill should be used and (2) how to execute the tasks. For PPA tasks, designers should develop intuition as to when to deploy each AI solution and how much it will impact the results to maximize return on resources used.

\begin{figure}[t]
  \centering
  \includegraphics[width=\linewidth]{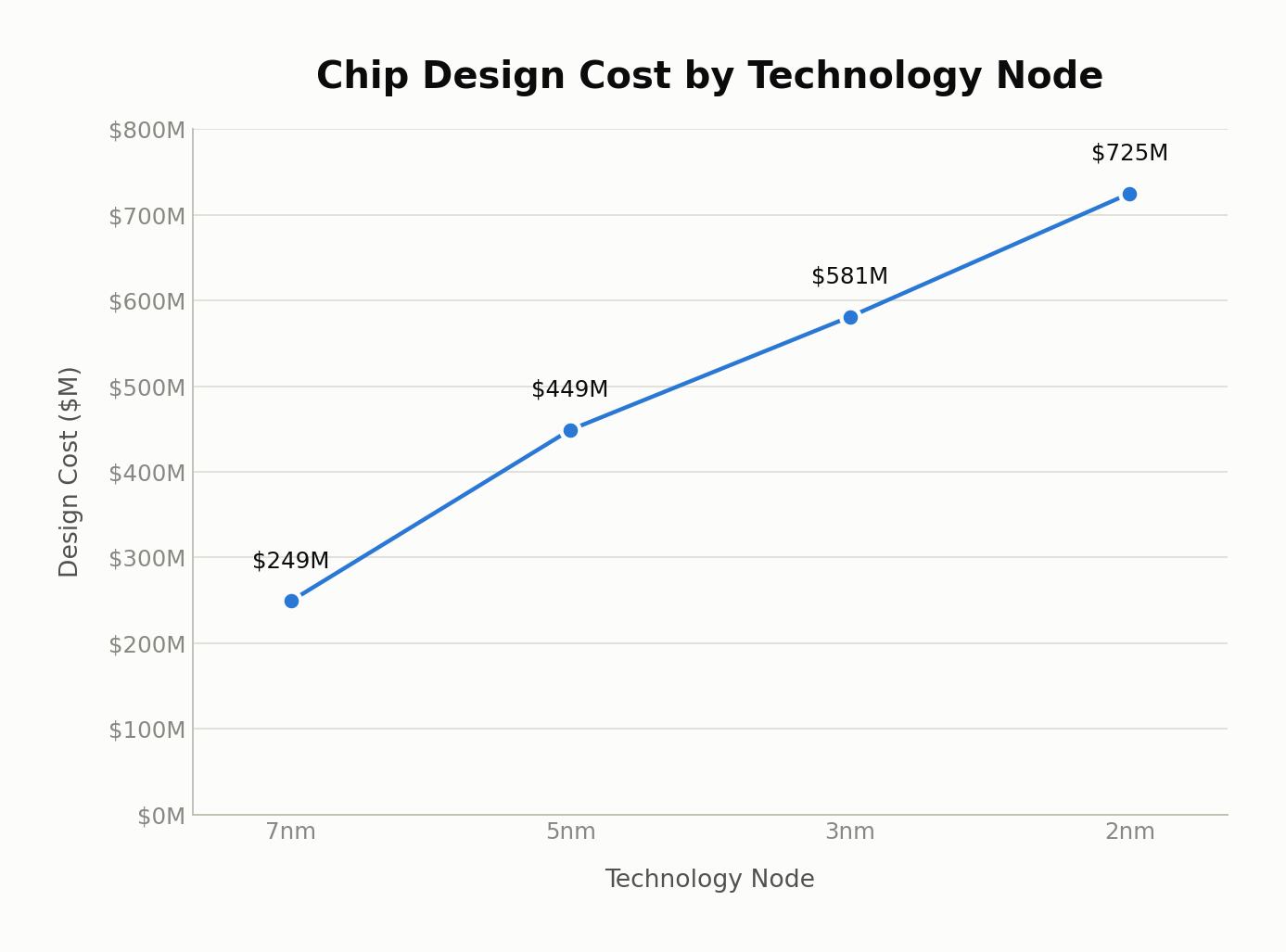}
  \caption{Projected chip design costs by technology
  node~\cite{tomshardware}. Modern technology processes incur substantially
  higher NREs which influence how we structure silicon design workflows.}
  \label{fig:design-cost}
\end{figure}

\smallskip \noindent \textbf{Resource Usage Intuition.} Silicon design already structures workflows and behavior around resource constraints. For instance, EDA/CAD tool licenses are typically charged per seat per year, so teams structure workflows to ensure that the license pool is efficiently shared to minimize resource usage. At a higher level, the resources required to develop modern chips are substantial (\autoref{fig:design-cost}) so we have strict sign-off and frontloaded validation checks to guard against steep losses.

Similarly, in an AI native era designers need to build intuition about the token cost efficiency in the context of our silicon workflows. This requires continuously checking and measuring token usage telemetry over time; this enables designers to correlate and build intuition as to how many tokens were used to achieve the work or impact produced, and audit for any surprises that defy expectations. Measuring token usage also enables designers to build intuition as to how to optimize token efficiency by tuning prompts, integrating more efficient deterministic tools, and comparing the token costs of different AI solutions. This token cost optimization and token efficiency intuition is especially important in light of the significant resource expenditures required to make AI native silicon development efficient.

\smallskip \noindent \textbf{Design Specification and Validation Intuition.} To harness AI effectively, silicon designers need to be able to craft precise and unambiguous problem specifications to express design intent. AI models currently lack expert silicon domain experience (due to data scarcity) so any ambiguity in a prompt specification increases the surface area for error (e.g., hallucinations, incorrect assumptions, etc.). Concepts that human engineers know reflexively such as timing exception rationale, power domain assumptions, design trade-offs, and clock domain crossing constraints are not always obvious to AI models. Thus, designers should build the intuition to understand what to specify in AI prompts so that design intent is fully encoded for AI tools. This AI intuition is analogous to how existing designers build highly precise IP specification documents and intuitively know what details are required to make the specification as unambiguous as possible (e.g., port interfaces, functionality, bit widths, protocols, etc.).

Subsequently, designers should develop the intuition and skills to be able to validate, trust, and explain whether AI-generated collateral is correct. To do this, designers should develop the intuition on how things can go wrong with AI-generated collateral and formulate appropriate checks, graders, and guardrails to enforce correctness. This is similar to how designers currently validate IP designs to catch corner cases and execution paths to ensure an IP design passes required functional checks before sign-off. AI-generated collateral will require the same type of intuition to also validate and guard against AI failure modes (e.g., solutions that look right but are functionally wrong). Finally, designers should develop the intuition and skills to understand and explain why an AI-generated design is correct; this requires new intuition and skills to be able to interpret and validate what the AI did so that it can be trusted for use and integration.

\subsection{Knowing When (Not) to Use AI}
\label{subsec:when-to-use-ai}

\begin{figure}[t]
  \centering
  \includegraphics[width=\linewidth]{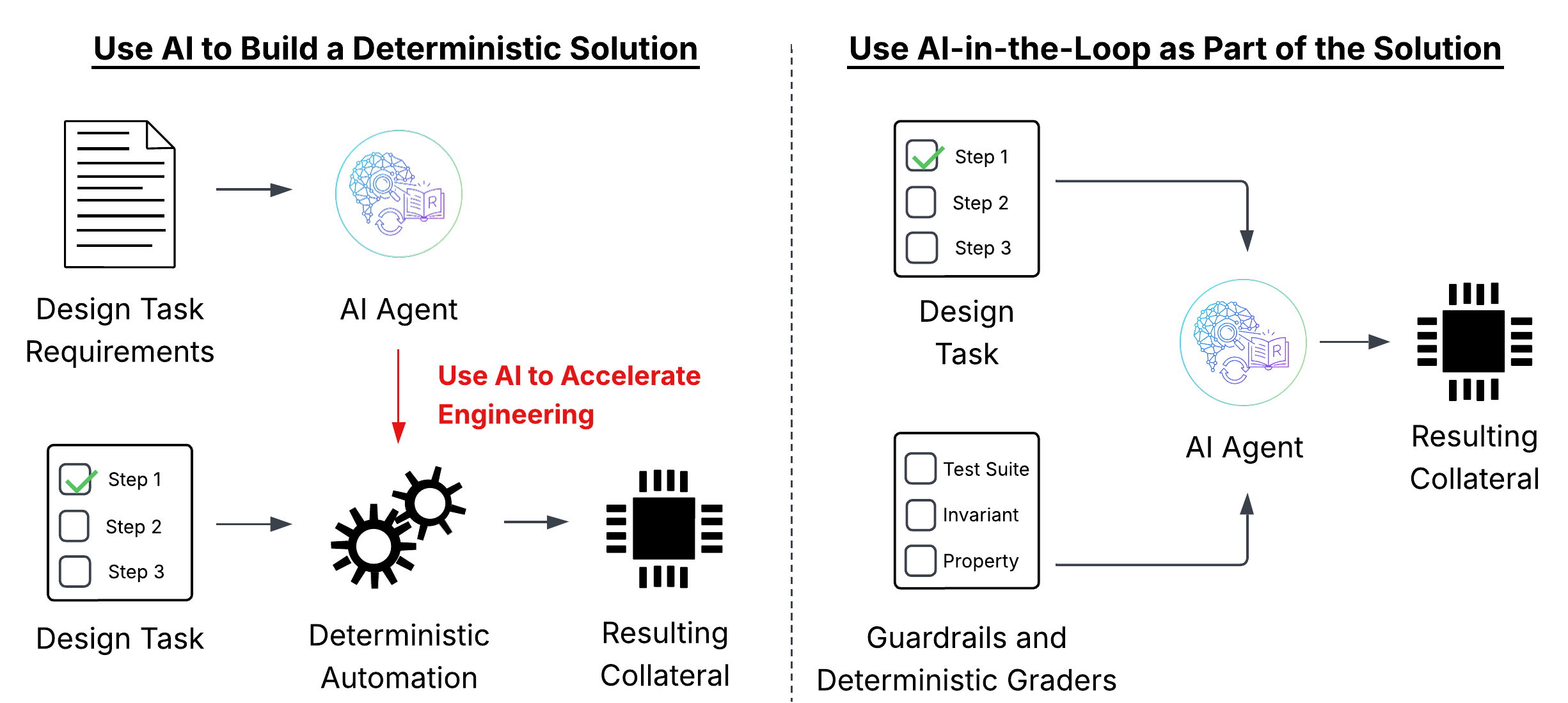}
  \caption{Using AI to build deterministic solutions is typically less resource intensive for known closed-form solutions. Using
  AI-in-the-loop is often more intensive and requires guardrails
  for problems where a closed-form solution may not be known.}
  \label{fig:when-to-use-ai}
\end{figure}

AI can be applied to accelerate the implementation of deterministic solutions, deployed in the loop to delegate tasks to AI agents, or both (\autoref{fig:when-to-use-ai}). Each approach has different tradeoffs which are worth building intuition about to ensure the most effective and efficient deployment of AI tools. Workflows which have AI-in-the-loop will continuously incur token costs which generally makes them more resource intensive than a comparable deterministic solution (if it exists). AI models also remain notoriously non-deterministic which requires (sometimes non-trivial) engineering guardrails to recover the guarantees of deterministic solutions. Deterministic non-AI solutions on the other hand tend to be faster and less resource intensive to operate but can only be implemented if a known closed-form solution exists and is sufficient.

To navigate this dichotomy, silicon designers should build intuition and expertise to distinguish when to use AI to build a closed-form deterministic solution (if it exists) and when it is appropriate to use AI-in-the-loop to build an agent. This intuition is typically acquired through academic training in silicon design fundamentals to discern what problems have deterministic closed-form solutions (e.g., polynomial time algorithms) versus those that are fundamentally difficult and may warrant AI-in-the-loop (e.g., spec-to-RTL lowering). Finally, AI-in-the-loop and deterministic processing can also be composed; however, deploying this hybrid approach again requires mastering the fundamentals to break down complex tasks to discern whether each task has a closed-form solution or may merit an agentic AI solution.

\subsection{Discerning Hard Problems and Emergent Capabilities}
\label{subsec:emergent-capabilities}

\begin{figure}[t]
  \centering
  \includegraphics[width=\linewidth]{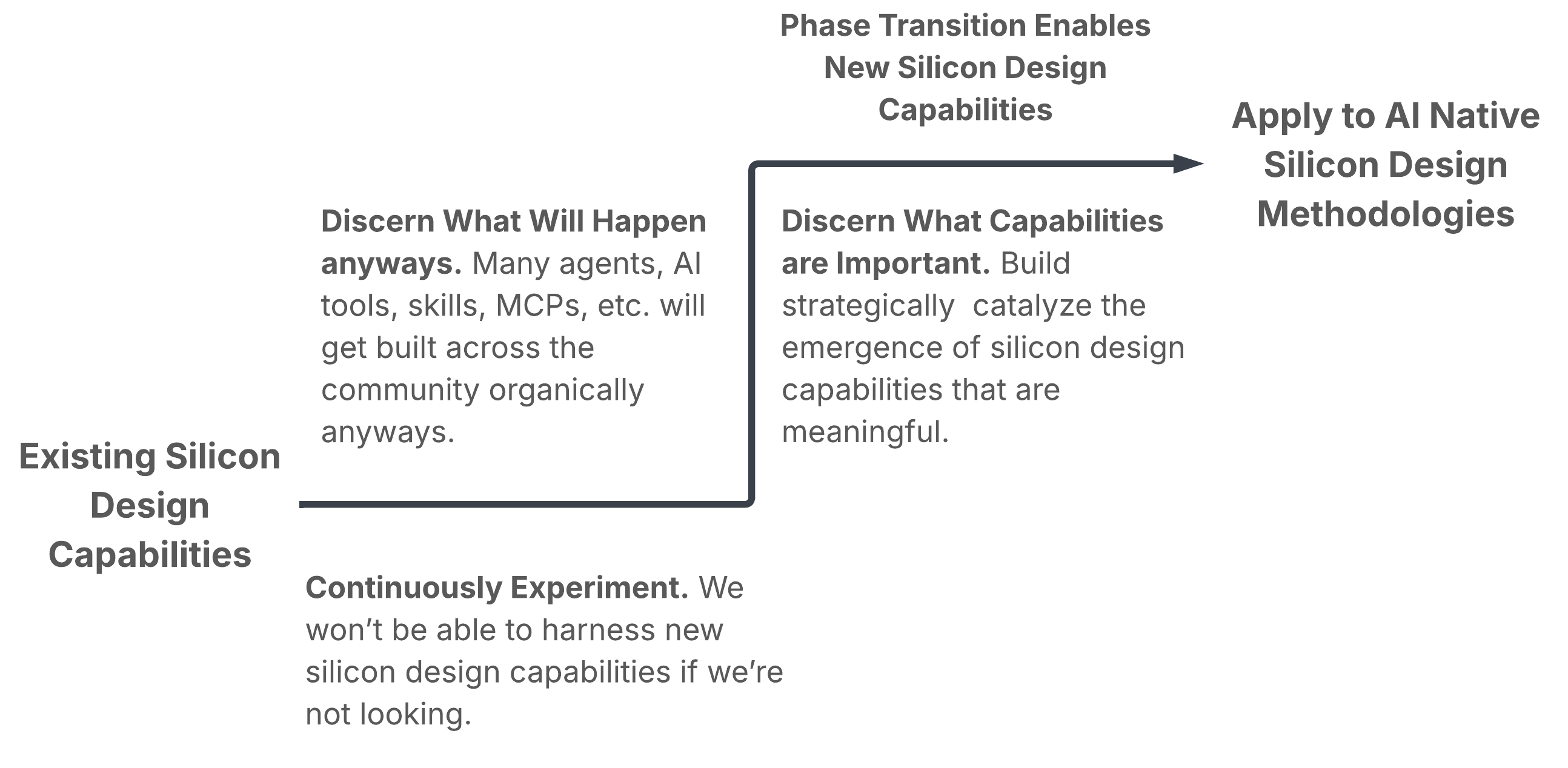}
  \caption{Many AI tools and agents will be built by the community
  organically. Practitioners will need to continuously experiment to watch for emergent new capabilities to harness for AI native silicon design.}
  \label{fig:emergent-capabilities}
\end{figure}

The AI silicon ecosystem and capabilities continue to develop and mature at an unprecedented rate. As a result, a wide variety of AI tools, agents, and integrations have emerged rapidly at scale (albeit in a somewhat chaotic fashion). Each individual AI tool or agent may only solve a small problem; however, over time many previously intractable silicon design challenges will suddenly become possible once a critical mass of AI agents and tools organically develop. To harness these powerful new capabilities, silicon practitioners should continuously monitor and experiment with emerging AI tools, agents, and integrations to watch for such emergent behavior (\autoref{fig:emergent-capabilities}). Similar to how Claude Code enabled powerful new disruptive software development capabilities, practitioners should watch for similarly emergent and disruptive capabilities for silicon design.

This requires mastering two core competencies: (1) continuously experimenting intentionally to monitor and identify when a capability phase shift occurs, and (2) recognizing which emergent capabilities are truly meaningful. Intentional experimentation relies on formulating clear hypotheses to determine when a novel, impactful capability has emerged within the AI silicon design ecosystem. This puts renewed importance on fundamental research training in academic institutions as a bedrock for lifelong experimentation in industry. Recognizing which emergent capabilities are genuinely transformative requires understanding where the remaining hard problems in silicon design lie. Without a firm understanding of what problems are difficult, such as timing closure complexity, analog-digital co-design, or reliability under process variation, designers cannot distinguish a true breakthrough from toy demonstrations.

\section{Tools}
\label{sec:tools}

Silicon EDA tools were originally developed for human operators but in the AI native era they will have to support both human and AI orchestrators which impacts how we interface and operate them.

\subsection{Electronic Design Automation in an AI Native Future}
\label{subsec:eda-ai-native}

AI tools for silicon design are being adopted across the industry at unprecedented speed. However, EDA tools, which are fundamental to silicon design, present a unique structural bottleneck. Unlike software engineering tools, traditional EDA tools are often internal, license-limited, and computationally intensive. EDA tools are typically slow since they provide the extreme precision and accuracy required to frontload design validation guarantees prior to sign-off since silicon cannot be patched or fixed after tapeout; this makes high-fidelity EDA tools indispensable and fundamental to silicon design. Unfortunately, the limited licenses create a fundamental scalability bottleneck since AI agents can be parallelized and instantiated limitlessly while EDA licenses cannot. The long runtimes also exacerbate the design iteration time when integrated on top of potentially long-running autonomous AI agents.

\begin{figure}[t]
  \centering
  \includegraphics[width=\linewidth]{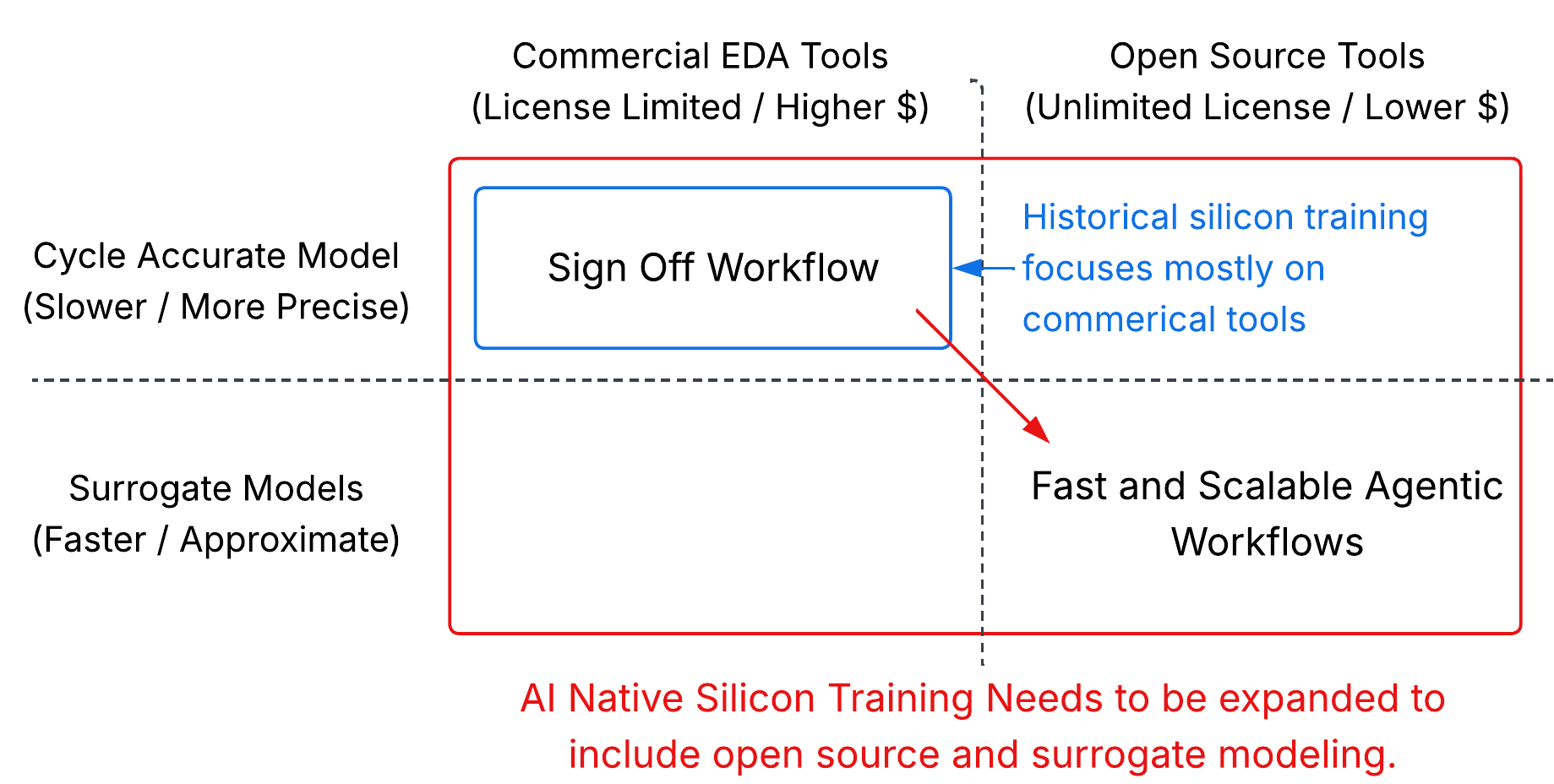}
  \caption{Open source tools reduce dependence on limited licenses while
  surrogate models trade accuracy for speed; these tools can enable rapid iteration at
  scale before invoking commercial sign-off tools.}
  \label{fig:tool-ecosystem}
\end{figure}

This presents a renewed opportunity for open source tools and surrogate models which trade accuracy for speed and scalability to bridge this gap and potentially opens the industry to a multi-tier EDA tool ecosystem (\autoref{fig:tool-ecosystem}). Open source tools cannot be used for sign-off quality checks but do not require licenses so they can be scaled limitlessly. Mid-level surrogate models on the other hand can serve as fast approximations which trade accuracy for speed to reduce the run time bottleneck. Together, these solutions can help alleviate the license limit and run time bottlenecks that need to be considered to scale agentic silicon workflows. We expect that commercial EDA tools will still be required for sign-off and final checks, but their role will remain limited by the number of licenses and run time speed.

To navigate this shift, designers should understand how and when to trade off execution speed and accuracy between traditional EDA tools and open source/surrogate models. All chip designs need to ultimately pass through sign-off tools for final validation, so mastery of licensed commercial tools will in the near term remain fundamental even in the AI native era. However, designers should also be able to discern when to opportunistically apply open source tools to complement silicon design workflows to unlock faster design iterations and productivity gains. As a result, AI native silicon practitioners should expand their fluency and familiarity to encompass both the commercial sign-off tools and open source EDA tools.

Silicon designers should also be able to evaluate when fast, approximate evaluations are reasonable and when only sign-off fidelity is acceptable. This requires a deep understanding of the fundamental mechanisms of different silicon design tasks to be able to discern when surrogate model approximations or open source tools can be safely deployed without compromising the technical rigor of design decisions and conclusions. For instance, high-level design space exploration may only require the relative ordering of design costs. On the other hand, microarchitecture design decisions for cache configurations or branch prediction may necessitate more precise, cycle-accurate granularity, while preparing a design for fabrication requires full sign-off EDA tool runs.

\subsection{AI Native EDA Tool Interfaces}

\begin{figure}[t]
  \centering
  \includegraphics[width=\linewidth]{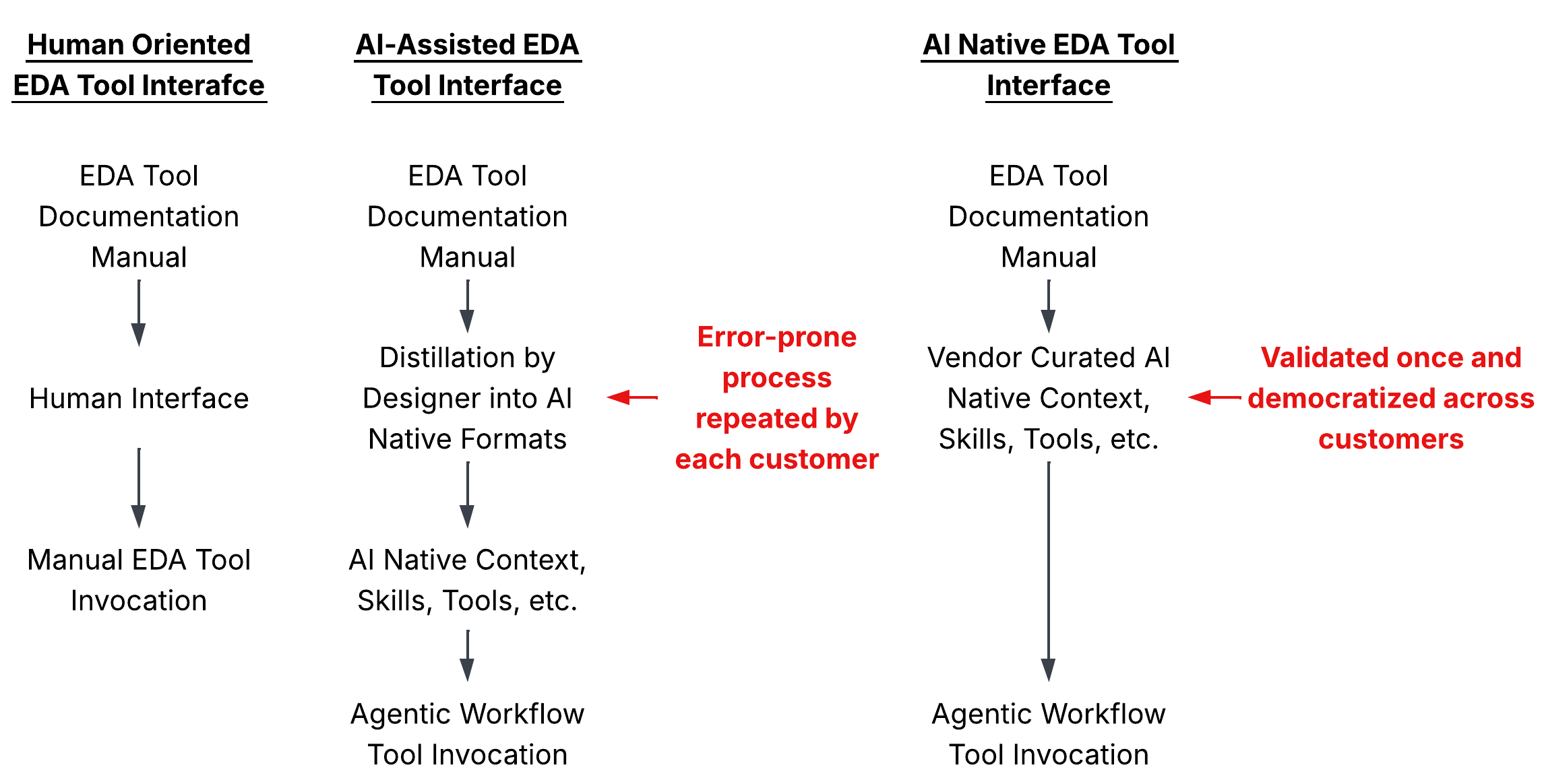}
  \caption{Human-oriented interfaces rely on technical documentation
  manuals. To bridge the gap, designers distill manuals into AI native
  formats (potentially error-prone). AI native interfaces will require
  vendor curated AI native formats that are validated once and democratized
  across customers.}
  \label{fig:ai-native-interfaces}
\end{figure}

In an AI native era, EDA tool interfaces will need to shift to support both human and AI agents instead of solely manually user-driven operation (\autoref{fig:ai-native-interfaces}). This is because an increasing number of EDA tool calls will come from AI agents or AI tool interfaces like model context protocols (MCPs). This means EDA tools will need to provide stable machine-readable interfaces that can be targeted precisely by AI agents to integrate and compose them within upstream automation and workflows. To do this, EDA tools will need to also ship with AI native infrastructure like skills, contexts, and plugin collaterals which streamline the interfacing and integration process with AI agents or workflows.

This is in contrast to the current model where tools ship with large monolithic documentation manuals which are intended for human use. Currently to bridge the gap, designers attempt to take human-readable documentation and use AI to decompose, organize, and distill them into AI native formats like context and skills. However, this process is inherently error-prone and difficult to validate since most users are not deep experts in the EDA tool design and detailed mechanics. Rather, EDA tool vendors should provide AI native formats where the curation and validation effort by the vendor can then be used to scale across customer teams. Without these established clean interfaces and curated precise AI native context and interfaces, it makes the EDA tool integration process into agentic workflows considerably less stable, more fragile, and more error-prone.

\subsection{Agentic Workflow Design for Silicon}
\label{subsec:agentic-workflow-design}

To build effective agentic workflows for silicon, practitioners should divide, conquer, and parallelize highly complex silicon design tasks. This then allows designers to modularize and reduce the size and complexity of each task delegated to either an AI agent or a deterministic tool. This reduces the complexity of the necessary prompt specification and the surface over which an AI agent can make mistakes. The finer granularity also makes the process of understanding and explaining what the AI did in each step more manageable to validate correctness. This auditability and explainability are again especially important for silicon design to make any collateral generated by AI trustworthy. Finally, the finer-grained agentic workflow design enables designers to specialize LLM configurations since certain LLMs may be better suited for certain design tasks over others.

At a higher level, modular agentic workflow design makes it easier to compose and reuse agents across different workflows (similar to software libraries). The modularity also exposes agent parallelism and potential runtime optimizations for highly complex agentic workflows to hide some of the potentially long EDA tool run times which enables potentially faster end-to-end design iteration time. Finally, as the complexity of agentic workflows grows, no single AI agent will likely perform optimally across every task in the silicon design flow. Without some hierarchical decomposition, the length of the specification and prompt required to fully specify the task will increase beyond practical human-managed limits and LLM context windows. As a result, we expect production workflows will likely shift towards orchestrating ensembles of heterogeneous agents, each specialized for different stages or sub-problems.

\subsection{Industry Standards and Interoperability}

\begin{figure}[t]
  \centering
  \includegraphics[width=\linewidth]{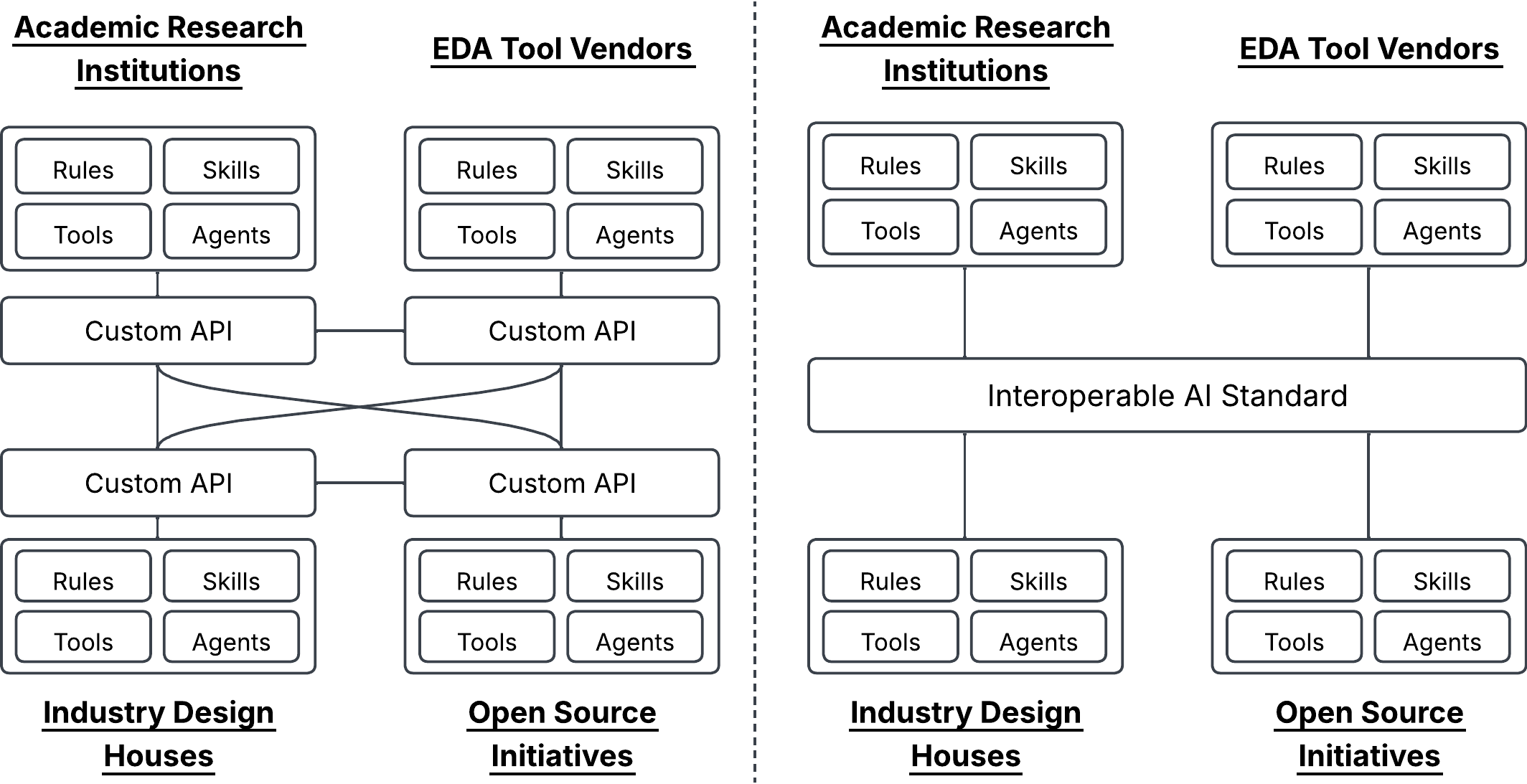}
  \caption{Without standardized interfaces, AI agents will need a
  quadratic number of custom adapters between institutions.
  Standardization enables seamless interoperability and portability by
  providing a common interface.}
  \label{fig:standardization}
\end{figure}

At an industry level, silicon design and productization will require coordination across industry entities like IP design houses, EDA vendors, academic institutions, and silicon fabrication companies (\autoref{fig:standardization}). This makes AI native standards and interoperability important since no single company can build the full evaluation suites and design flows due to the size of the design space, range of technology nodes, and proprietary nature of the designs, EDA tools, and flows. If formulated properly, standards and interoperability APIs can abstract away internal details but still enable AI native design. To do this, the industry will need to develop standards and interoperability mechanisms via joint initiatives like Si2 LBC~\cite{Si2, LBC} to curate the shared standardized benchmark suites for AI agents and infrastructure. For example, this may include releasing reference designs to ground agent evaluations and evaluation harnesses for the ecosystem to train and validate against. This is analogous to how MLPerf~\cite{mlperf} created a standardized and interoperable benchmark and evaluation platform for comparing ML hardware performance.

Standardized interfaces will also be important to enable evaluation of different competing agents for a silicon design task. For instance, an IP design team looking for a specification-to-RTL agent may have to choose from several different agents which have different tradeoffs in operating costs, run times, or PPA results. To effectively compare and contrast competing solutions, standardized interfaces may be necessary to control for as many configuration and operating variables as possible to ensure evaluations are fair. Without compatible interfaces, comparing and explaining the strengths and limitations of competing solutions becomes significantly more difficult.

Standardized interfaces also enable seamless agent composition and interchangeability. In a mature ecosystem, agents may be supplied by different vendors and combined with either first-party or other third-party agents. To enable agentic capabilities like agent-to-agent communication, agents need well-defined schemas so that different agents can hand off results to one another without bespoke (and potentially error-prone) integration work. Standardized interfaces also allow workflows to seamlessly interchange agents if a better competing solution is made available. Otherwise, each agent integration risks becoming a custom engineering effort that slows down agentic workflow development. Since the landscape is still shifting, the exact standards will likely take some time to converge, but designers need to understand and monitor the latest state of the art for making agents interoperable.

Finally, standardized interfaces and interoperability also provide a way to formalize what guarantees or issues an agent returns. For instance, if an agent generates collateral which passes lint and CDC checks, that guarantee (and any proof) is valuable to pass to any downstream agent to avoid having to re-execute checks. Conversely, if an agent encounters a failure or produces low-confidence results, a standardized interface to communicate the nature of the error, confidence, debug information, logs, and telemetry is important to trigger appropriate escalation policies and workflows to avoid letting failures or low-quality results silently propagate through the flow.

\section{Adapting Academic Pedagogy for AI Native Silicon Design}
\label{sec:pedagogy}

\subsection{Using AI as a Learning Aid}

AI tools and fluency have vastly expanded the knowledge a silicon designer needs to master in order to operate at the cutting edge of silicon design flows. In other words, on top of the foundational first principles training, additional AI skills, training, and experience are now needed to practice advanced silicon design. This does not mean that yesterday's core silicon principles are no longer relevant; rather, it raises the bar for the amount of training students need to master.

To accelerate and compress the volume of training in the education process, it is worth considering how to structure AI tools to supplement the learning process while guarding against excessive cognitive offload. When used correctly, AI serves as a powerful accelerator that deepens a student’s grasp of core concepts and functions as a learning aid rather than a substitute for actual learning. We are already seeing students teach themselves using AI to accelerate the rate at which they study material, facilitate absorption of deeply technical knowledge, and become domain experts in highly arcane subjects. To successfully manage and compress the training required for AI native practice, students should be encouraged to view these tools as a way to augment their intelligence and learning process rather than another exercise in cognitive offload.

Ultimately, fostering the proper adoption of AI to build genuine understanding requires each student and practitioner to be honest with themselves about whether they are building true mastery or merely satisfying surface-level requirements. Educators can teach the basic frameworks and approaches to how to harness AI effectively but at the end of the day it is up to the students to apply and put it into practice and hold themselves accountable. Similarly, for advanced research, AI can be a powerful aid if used properly, to meaningfully complement and move the research frontier forward; otherwise, it risks serving as a substitute for fundamentals which limits the foundation on which advanced research can be built.

\subsection{Adapting Pedagogical Strategies}

To adapt academic training for an AI-native world, pedagogical strategies will need to realign incentives so students master foundational principles before relying on AI automation. One concrete way to do this is restructuring assessment weights to manage student incentives directly. Educators can establish an environment where foundational mastery is non-negotiable while still acknowledging the need to learn AI skill sets by capping AI-enabled homework and lab assignments at a small fraction (e.g., 10–20\%) of the overall course grade while weighting proctored, closed-device exams heavily (e.g., 80–90\%). The rebalancing of incentives encourages students to build fundamental, manual problem-solving skills, while the lower-weighted homework still incentivizes experimentation to build AI fluency. There are also proposals to go back to bluebook exams or labs without AI tools to more strictly control the learning environment but how they are used in academic practice will likely depend on the course, institution, and education philosophy.

In addition, assignments will need to encourage students to actively verify, question, and critically evaluate AI results rather than passively generating output. For instance, in hardware engineering a prompt might require students to generate an INT8 multiply-accumulate (MAC) unit in Verilog using AI, but the core grade relies on their ability to reconcile that AI output against classroom theory such as explaining orally specifically why the design is or is not optimal. Similarly, instructors can provide a rigorous ``golden testbench'' with demanding pass criteria where students debug and explain why AI-generated HDL failed, or conversely, have students hand-craft testbenches to validate a provided device under test (DUT). Crucially, generating both the DUT and the testbench using AI needs to be strictly discouraged; fully automated loops strip away the essential validation and learning feedback loop.

Finally, pedagogical frameworks need to integrate training for soft skills to synthesize, justify, and communicate results ideally through direct human evaluation. Incorporating oral exams provides a powerful mechanism to probe a student's genuine depth of understanding, forcing them to articulate why specific design choices were made rather than relying on functional code alone. For large course enrollments, this evaluation can be broken down and delegated to lab or group project settings to make the human evaluation process more scalable. Since these soft skills can take time to develop, it will be important to consistently integrate these human evaluation touch points throughout the curricula so that students can hone them sufficiently prior to entering the industry.

\subsection{Evaluating Students for Industry Practice}

As students enter the workforce, academic and technical evaluation methods will need to adapt to also assess AI native fluencies. We are already seeing instances where industry interviews now allow for the use of AI tools~\cite{meta_interview, google_interview} but it is still an open question as to what the best approach, interview formats, and technical questions are to properly assess candidates. However, what likely will not change is the need to evaluate and critically think through problems based on core fundamentals and formulate complementary AI techniques. As a result, the industry evaluation process will need to put renewed (but not excessive) emphasis on shifting how we evaluate silicon design fundamentals while also accounting for shifts in how AI changes how we practice silicon design.

One of the most direct ways this shift will manifest is in technical interviews which will need to adapt to gauge both AI fluency and technical fundamentals simultaneously. For example, an interviewer might ask: “Given a task X (for example, software optimization for vector units), how would you develop an AI-assisted solution? Are there places where you would not use one, and why?” This type of problem provides candidates an opportunity to demonstrate how to precisely divide and conquer the problem as opposed to blindly applying AI. It also mixes a truly difficult problem with problems where known (partial) solutions may exist and provides the candidate an opportunity to demonstrate how well they can discern between when (not) to use AI. We expect that similar types of evaluations will need to be crafted to gauge other AI native competencies as outlined earlier in this work as we move into the AI native era.

\section{Technical Training for AI Native Silicon Design}
\label{sec:training}

The skills and best practices for designing agentic systems are complementary to existing core principles and fundamental training; thus academic training will need to both reemphasize historical core training courses and integrate AI fluency into curricula.
\autoref{tab:training} shows how core conventional academic training is still relevant to the agentic thinking for industry practice today. The exact course structures are different across training programs so the precise interpretations will vary by institution.\footnote{The technical training here does not diminish the relevance of other existing courses.}

\begin{table*}[t]
  \centering
  \caption{A list of technical training courses, the key fundamental skills
  that they cultivate, and their relevance to the AI native silicon design
  future.}
  \label{tab:training}
  \small
  \begin{tabular}{@{}>{\raggedright\arraybackslash}p{0.22\textwidth}
                    >{\raggedright\arraybackslash}p{0.34\textwidth}
                    >{\raggedright\arraybackslash}p{0.38\textwidth}@{}}
    \toprule
    \textbf{Relevant Training}
    & \textbf{Key Fundamentals}
    & \textbf{Relevance to AI Native Silicon Design} \\
    \midrule
    Software Engineering Principles for Silicon Design
    & Software Engineering Best Practices \newline
      Building Hardware like Software \newline
      Designing and Implementing Complex Systems
    & Provides foundational software basics and best practices that AI
      tools are built upon and assume \\
    \addlinespace
    Silicon Design Basics
    & Analog Mixed Signal Circuits, Physical Design, Circuit Theory, PPA Optimization,
      Simulation and Modeling, DV, DD, DFT, EDA, SoC, Architecture
    & Core Fundamental Silicon Design Principles and Basics \\
    \addlinespace
    Algorithms
    & Divide-and-Conquer \newline
      Complexity Analysis \newline
      Algorithm Design
    & Problem Decomposition and Mapping to Agentic Workflows \newline
      Discerning Hard Problems and When (Not) to Use AI \newline
      Recognizing Emergent Design Capabilities \\
    \addlinespace
    Design Validation and Verification
    & Design Implementation Methodology \newline
      Testing and Validation
    & Prompt Specification Engineering \newline
      Encoding Design Intent \newline
      Formulating Deterministic Guardrails and Graders for AI Agents \\
    \addlinespace
    Technical Communication
    & Expressing and Specifying Design Intent \newline
      Communicating Results Efficiently
    & Prompt Engineering and Fully Specifying Design Tasks \newline
      Design Decision Justification \newline
      Explaining and Justifying AI-Generated Results \\
    \addlinespace
    Computer Architecture and Systems Design
    & Building, Testing, and Validating Large-Scale Silicon Systems
    & Designing and Testing Large-Scale Agentic Silicon Workflows \\
    \addlinespace
    Random Processes
    & Reasoning about and Designing with Randomized Behavior
    & Designing Tools (i.e., agentic workflows) with Randomized Processes in
      the Loop \\
    \addlinespace
    Advanced Silicon Research Topics
    & Continuous Experimentation \newline
      Advanced Research Methodology \newline
      Open Source EDA Tool Fluency
    & Watching for Emergent AI Enabled Capabilities \newline
      Formulating Hybrid OSS/Sign-off EDA Tool Agentic Workflows \\
    \bottomrule
  \end{tabular}
\end{table*}

\smallskip \noindent \textbf{Software Engineering Principles.} Agentic systems and automation will be backed by software (even if it is AI-generated). Thus, basic training and mastery in software engineering principles and best practices are necessary even for AI native silicon:
\begin{itemize}
\item Basic Software Design Principles. These include fundamentals like good API design, inheritance, abstraction, hierarchical organization, reusable libraries, etc.
\item Version Control and Modular Commits. Autonomous AI agents will submit changes alongside human workflows. To minimize conflicts, silicon designers should use version control and modularly create commits/pull requests to operate alongside AI.
\item Modular Design and Reuse. To compose and reuse agents and workflows, agents need to be built modularly with standardized interfaces. Building reusable components also reduces how much code needs to be written (potentially by AI) and revalidated.
\item Unit and Integration Testing. To move at the speed of AI, designers need to use tests and harness automated regression testing infrastructure to preserve correctness. AI can repair or fix code automatically but only if it knows what the intended correct behavior is via tests.
\item Reproducibility. For agentic workflows, individual components need to be stable so results are reproducible. Otherwise, it becomes much harder to debug when things go wrong.
\item Codebase Organization. Complex agentic systems mean silicon designers need to be able to manage a complex codebase. Large monolithic codebases quickly become unmaintainable; modular organization helps manage this complexity.
\end{itemize}

Ultimately AI tools are built upon software engineering best practices; thus to layer silicon design on top of AI and get the most out of AI capabilities, silicon practitioners will have to learn a minimal amount of software engineering principles.

\smallskip \noindent \textbf{Algorithm Design and Complexity Analysis.} Agentic workflow design requires strong foundations in algorithms and complexity analysis which provide the mental frameworks and skills to: (1) divide and conquer complex problems, (2) discern what problems are genuinely difficult, and (3) understand when algorithmic approximations are appropriate. Divide and conquer is relevant to agentic workflow design because it enables designers to break down and partition monolithic, highly complex problems into more manageable tasks. These tasks can then be delegated to specialized agents and it is easier to fully specify the task for the AI (see \autoref{subsec:agentic-workflow-design}). The finer-grained problem decomposition also enables designers to map whether a task can be delegated to a deterministic tool or an AI agent depending on the difficulty of the subproblem (see \autoref{subsec:when-to-use-ai} and \autoref{subsec:emergent-capabilities}). Formal complexity analysis provides the foundational skills to conduct this analysis to ensure that applying AI is warranted and avoid costly AI overengineering for simple problems which may have deterministic and less resource intensive solutions (see \autoref{subsec:classical-intuitions}). Finally, the algorithm coursework provides the foundational skills to apply and leverage algorithmic approximations which directly translate to reasoning about when and how to leverage surrogate model approximations (see \autoref{subsec:eda-ai-native}).

\smallskip \noindent \textbf{Digital Design Specification and Validation.} As silicon workflows shift from “how do I implement the design” to “how do I specify, bound, and judge the work done by AI” (\autoref{subsec:validating-ai-collateral} and \autoref{subsec:classical-intuitions}), technical training in silicon design validation will become more important. To effectively prompt AI agents, practitioners need to learn how to write complete and precise specifications to express design intent. This training is not unlike skills learned in digital design courses where students learn to formulate and engineer against precise IP specifications and ensure that the design validates against expected specifications. These courses also provide the training engineers need to discern edge cases and anticipate non-obvious failure modes to ensure specifications are as complete as possible. These skills translate to the technical toolkit needed to construct complete specification prompts as well as to formulate the deterministic validation, grader, or guardrail checks required for agentic workflows. Because so many things can go wrong, it will be important for silicon designers to encounter the various types of failure modes in practical training at least once to understand what they need to watch out for.

\smallskip \noindent \textbf{Technical Communication and Soft Skills.} AI will gradually shift workflow bottlenecks towards critical thinking, communication, and other collaboration workflow aspects (see \autoref{subsec:workflow-bottlenecks}). This puts renewed importance on fundamental ``soft'' skills such as coordination and technical communication with teammates, upper management, and third parties. Soft skills are particularly important in practice since there are more collaborators and industry processes to negotiate before solutions get adopted or deployed. Once a solution is complete, designers need to be able to articulate and communicate what they built and why it should be adopted or deployed to collaborators. It is not uncommon for designers to come up with a technical solution only for it to never see deployment because it was never communicated, demonstrated, or brought up for proper review. Thus, even in an AI native era where implementation can be very fast, practitioners will still need to have the skills to convince collaborators to use and deploy their solutions.

\smallskip \noindent \textbf{Computer Architecture and Systems Design.} Designing, implementing, and debugging autonomous silicon agent workflows requires a systems-level perspective, making academic coursework in computer architecture and systems design essential in the AI era. As a designer's role shifts towards orchestration, the primary engineering challenge shifts from lower-level implementation and execution logic to managing, orchestrating, and debugging complex systems of agents to design entire systems. Computer architecture coursework teaches students to manage holistic system complexity at this higher level of abstraction. Furthermore, hands-on validation of large silicon systems equips engineers with directly transferable skills for complex agentic workflows such as exposing appropriate telemetry to direct debugging, diagnosing and fixing intermittent failures, and pinpointing and fixing where things go wrong at scale.

\smallskip \noindent \textbf{Randomized Algorithms.} Building AI agents requires reasoning about, designing around, and evaluating under non-deterministic black-box LLM behavior. LLMs are effectively probabilistic engines which lack deterministic guarantees unless there are guardrails to validate the result. Thus, designing agentic systems shares many parallels with designing algorithms around stochastic components like simulated annealing, genetic algorithms, etc. Similarly, designers should understand the basic mechanics of randomized algorithm design which directly transfer to how to reason about the mechanics of implementing, evaluating, testing, and debugging non-deterministic agentic workflows. This training also builds intuition as to where things can go wrong and how to design guardrails which transfer directly to agentic workflow design challenges.

\smallskip \noindent \textbf{Continuous Experimentation.} Silicon practitioners in the near term should continuously experiment to keep up with the speed at which AI is currently maturing and for emergent design capabilities. This requires equipping practitioners with a lifelong continuous experimentation mindset and tools to watch for and understand when truly transformational silicon design capabilities emerge (see \autoref{subsec:emergent-capabilities}). Building this skill set in academic training usually requires instruction on basic research principles and methodology (i.e., scientific method). This continuous experimentation mindset can take many forms from continuously trying to solve a problem with the latest LLMs, mixing and matching different agents to evaluate their capabilities, or setting up large-scale, continuously running experiments to explore what is possible. Building the skill set to set up, execute, and interpret experimental results will be increasingly important for AI since the complexity and scale of these systems make it increasingly difficult to draw airtight empirical data-driven insights and conclusions.

\smallskip \noindent \textbf{Hands-on Experience.} Even with AI, there is still no substitute for hands-on technical experience to understand all of the technical nuances, edge cases, failure modes, and soft skills that go into building silicon. This is because industry workflows and design problems tend not to fit perfectly into purely theoretical textbook training frameworks and pedagogy. To bridge this gap, we typically use intentionally designed hands-on lab training, group projects, and assignments to enable students to gain direct technical experience to build these technical skills. This builds familiarity with the realities of the design process such as working with EDA tools, evaluating synthesis reports, debugging complex verification environments, and measuring corner-case coverage which can then be transferred to an industry setting. Training on how to use AI tools for silicon will be no different and will require a similar degree of hands-on training with AI tools to enable students to build familiarity and intuition which later can be applied to industry practice (see \autoref{sec:intuition}). Finally, we note that in practice there is wide variability in AI tools or agents, token constraints, and EDA tool flows so students will need to learn transferable intuition and skills between tools, versions, etc.

\section{AI Native Advanced Silicon Research}
\label{sec:research}

This section provides some high-level views on how we expect AI to give rise to new advanced silicon research and development methodologies.

\subsection{AI to Accelerate Research Scaffolding}

Today, research students often complain that chip tapeout or large-scale research projects consist mostly of engineering work (e.g., physical design, verification, integration, and toolflow orchestration) and only a small amount of research. With AI, there is an opportunity to automate more of the engineering work to rebalance the time students are able to focus on research and innovation. To do this, students need to understand how to accelerate construction of the base scaffolding and boilerplate required to support their desired research workflows and prevent knowledge loss across generations of researchers. This requires students to learn how the tool flows operate, where they fail, and how design decisions propagate through the stack before they can effectively direct, diagnose, or validate AI-generated implementations; it will also require building similar AI native intelligence pools that can be transferred across generations of researchers to reduce boilerplate or reconstruction of knowledge bases. Without a rigorous understanding of the basics of how the flows are formulated, students will not be able to specify and orchestrate AI agents to accelerate implementation, or capture and encode the expertise to facilitate more efficient research. To encourage the absorption of these fundamentals, educational curricula for silicon design will need to be staged where students first develop expertise in the complete chip-design process, and then use AI to learn how to accelerate implementation of these flows as opposed to blindly asking the AI.

\subsection{Revisiting Old Research Problems}

Advanced agentic AI has the potential to breathe new life into mature areas of research that have enjoyed decades of investment as well as emerging technologies and methodologies which were ahead of their time. Agentic AI can be combined with virtually any research field so it can quickly become overwhelming to identify the most promising returns on investment. To better understand the potential of individual research directions, researchers should first understand fundamentally why a research problem was originally difficult and whether advanced agentic capabilities change the calculus. Several key ways (among others) that agentic AI can make it worth revisiting a research field are:
\begin{itemize}
\item Commoditized Engineering. A research direction may historically have plateaued due to the prohibitively high engineering investment required to make progress, or the lack of a reasonable return on investment; since AI commoditizes engineering, these directions are likely worth revisiting.
\item Non-Human Interpretable Complexity. A research direction may require insight that defies human logic or is too complex for humans to reason through. Since AI can reason over more complicated patterns, revisiting these directions can unlock novel optimizations or solutions.
\item Cross-disciplinary Expertise Requirements. A research direction may require one (or more) different domains of expertise limiting how much an individual researcher could push the science. AI superintelligence can reduce the barrier to bridge these gaps in cross-disciplinary expertise.
\end{itemize}

Examples of research problems in the silicon space that can benefit from commoditized engineering are areas such as large-scale systems design, architectural benchmarking, or novel compiler design. These are areas which require significant amounts of engineering to even get started. On the other hand, research in emerging computing areas like asynchronous circuits, alternative memory technologies, stochastic computing, and analog-inspired circuits can benefit from both the commoditized engineering capabilities and agentic AI’s ability to reason about arcane compute encodings and design problems. Ultimately, it will require individual researchers to interpret and think critically about how AI changes the historical bottlenecks for their individual research interests to capitalize efficiently on the new research opportunities unlocked by agentic AI.

\subsection{Adapting Intellectual Research and Development Culture}

In the age of AI, what defines a good PhD student, professor, or research practitioner still begins with core fundamental qualities which have not changed. Rather than replace human intellect, AI serves as an advanced tool in the researcher's toolbox to accelerate research discovery. Researchers still have to possess the foundational capability to interpret results, formulate rigorous experimental methodologies, and explain their findings without relying on AI. Even with AI, researchers should be able to apply the core scientific method principles rigorously to validate and produce high quality research. For PhD students, the doctoral journey remains a vital training ground for acquiring and mastering these technical fundamentals and research aptitudes for careers in academia or industry. These programs also still remain the primary training grounds to hone essential soft skills such as taking initiative, communicating complex ideas, and defending technical decisions.

More generally, the broader silicon research culture needs to evolve in terms of how innovation is communicated, evaluated, and built upon. Traditionally, technical papers were the primary vehicle for documenting innovation, transferring technical insights, reproducing results, and cumulatively building upon them. However, as AI increases the sheer volume and velocity of published papers, this traditional communication model no longer scales to match the research throughput. To adapt, researchers should rethink how we collaborate, make results reproducible, and share knowledge at the speed of AI. Unless research is fully reproducible through shared code or detailed methodologies that allow others to build upon it, it will not deliver the full potential value to the scientific community.

Finally, keeping academic silicon research attuned with real-world practice requirements cannot be accomplished by academic research institutions in isolation. Closing this gap will demand tighter, more active collaborations between industry practice and academic institutions such as joint forums, guest lectures, invited talks, and shared research initiatives. These mechanisms and forums allow industry leaders to provide continuous visibility into emerging practices, tools, and talent gaps which will be important towards maintaining the talent pipeline for the industry.

\subsection{AI Native Research Technology Transfer Model}

\begin{figure*}[t]
  \centering
  \includegraphics[width=0.8\textwidth]{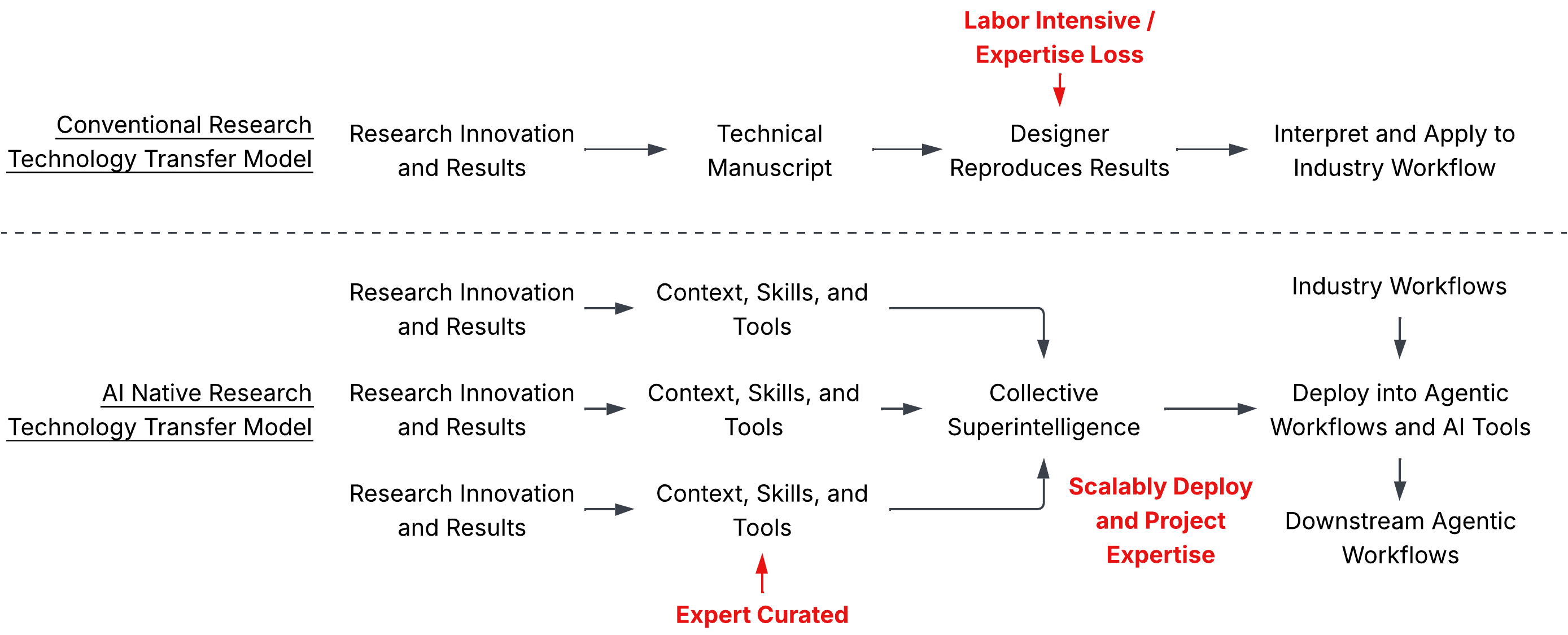}
  \caption{Conventional research technology transfer is labor
  intensive and lossy. AI native research technology transfer can
  enable more frictionless scalable deployment of advanced research
  techniques.}
  \label{fig:ai-native-tech-xfer}
\end{figure*}

One way to adapt how we communicate and transfer research is by adopting a more AI native technology transfer model to enable industry to better absorb and integrate the increasing volume of silicon research. In the past, dense technical text and (often) long technical formats were necessary for technology transfer since the intended audience was a human expert. However, in the AI native era this is gradually changing and it is becoming increasingly important to curate technical innovation for both humans and AI agents. To support this transition, researchers should adapt how to communicate and transfer advanced research to be more AI-compatible (see \autoref{fig:ai-native-tech-xfer}).

The research community is already starting to see shifts and proposals towards AI native mechanisms like agent-native research artifacts~\cite{agent_native_artifacts}, Claude Code plugins, and skills which enable pathways towards making technical contributions directly deployable with advanced AI agents and workflows. By packaging research innovations into AI native formats (i.e., context, skills, tools, agents, etc.), researchers and industry practitioners would be able to quickly transfer and deploy these innovations at scale.

Once advanced technical research and innovation are encapsulated into these AI-native-compatible representations (i.e., rules, skills, tools, agents, etc.), they can then be scalably transferred across academic and industry settings. This then allows the industry to leverage advanced AI’s ability to amplify engineering expertise via collective superintelligence to deploy these innovations by simply importing these assets. This is in contrast to the historical research and technology transfer model which required manual review and reproduction of research results which does not scale beyond a few papers and limits the impact to a few invested teams.

Silicon design research should still need resolve how to expose EDA tools (potentially with license restrictions) for reproducing power, performance, and area results to fully realize this new AI native technical research model. Open source tools that are not limited by licenses and intellectual property restrictions are likely one part of the solution but (at the time of writing) the details of the research methodology are still an open question. Thus, going forward the research community should work through these technical details as the industry moves forward and potentially consider merging AI native research formats into reproducibility evaluation processes such as artifact evaluation or technical paper submissions to facilitate technology transfers.

Finally, this AI native technology transfer model also generalizes to collaborations between research institutions and beyond silicon design. For researchers, this means that it provides an opportunity to assimilate and harness advanced research and superintelligence in fields that they are not native experts in. This is particularly important for conducting interdisciplinary research or quickly becoming deep experts by using these advanced research knowledge bases and intelligence. For instance, a computer architecture researcher may leverage superintelligence contexts about sustainability to be able to conduct sustainability-aware silicon design, or a chief architect may be able to reach further into the algorithm design to explore codesign opportunities by becoming an expert in that particular algorithm.

\subsection{Opportunities for Reinforcement Learning}

The chip design process is not a single decision but the result of thousands of design decisions from architectural choices down to placement and timing optimization. Since no decision is made in isolation, the merit of each design needs to be evaluated in the context of all subsequent decisions in the flow. The collective quality of these decisions is only apparent after the full flow completes, which creates a delayed reward signal. As a result, one way to cast the silicon design process is as a reinforcement learning (RL) problem where we need to optimize sequential decision-making under uncertainty and delayed reward. To capitalize on the theory, it is potentially worth revisiting how to structure and design AI workflows within this higher-level RL + agentic AI framework.

An RL agent generates a candidate design or modification, evaluates it against a grader for correctness, uses efficient checks for rapid exploration and resources-intensive sign-off tools for validation, updates its policy based on the reward signal, and repeats. This process creates its own training data since each iteration produces a (state, action, reward) tuple that the agent can learn from. This avoids the data scarcity challenge since operating the RL formulation only needs a well-defined evaluation function to deterministically grade results. The concrete methodology to apply this to silicon design is still an open question but it provides one opportunity forward towards enabling agentic silicon design despite the lack of silicon data for LLMs.

\subsection{Discerning What Will Happen Anyway}

Research resources are still finite and even with research agents there are compute capacity limitations so researchers should still operate efficiently in the age of AI. However, with the size and speed at which the research landscape changes, it can be overwhelming for individual researchers to understand what to work on. To make the research investment process more efficient, researchers should be able to discern what AI tools, capabilities, and innovations will happen organically. Reusing capabilities, tools, and components that already likely will exist reduces the amount of redundant work so that researchers can focus on solving the longer-term hard problems. Otherwise, research workstreams risk rediscovering the same innovations in parallel which dilutes the overall return on research investments.

In addition, because many competing solutions for the same problem will likely emerge, it is important for researchers to be open to collaborating and combining different approaches since the precise technical approaches are unlikely to be the same (at least in the near term). For instance, multiple parallel efforts may implement Karpathy loops~\cite{karpathy_loop} to optimize the same type of program but the underlying skills, strategies, and prompts may differ. The research community should embrace this dynamic and be open to collaborating to ensure these innovations can be properly reproduced, composed, and made complementary to enable technical solutions that are greater than their individual components.

\section{Conclusion}
\label{sec:conclusion}

There is no question that AI will transform how we design silicon and the industry will remain forever changed in the AI native era. In this work, we provided a view and discussion of how academia and industry can add structure to how to navigate this transition. We note that the perspectives and positions articulated here represent a snapshot in time; while we attempted not to make overly prescriptive recommendations, the positions in this work will need to be periodically revisited as silicon practice continues to evolve. Ultimately, the transition to AI native silicon design will likely be a democratized process which requires collective and joint efforts from across the industry and is larger than any individual institution. As a result, it will be important for the community to continue to exchange views of the challenges and potential solutions to jointly ensure that academic training and industry practice remain coupled to prepare the next generation of scientists and engineers for an AI native future.

\bibliographystyle{plain}
\bibliography{references}

\end{document}